\documentclass{subdwarf2}
\usepackage{graphicx}
\usepackage{tabularx}
\usepackage{longtable}
\usepackage{rotating}

\received{  }
\revised{}
\accepted{}
\submitjournal{ }

\shorttitle{Search for hot subdwarf stars in Gaia DR2 }
\shortauthors{Lei et al.}

\begin{document}

\title{New hot subdwarf stars identified in Gaia DR2 with LAMOST DR5 spectra }

\correspondingauthor{Gang Zhao}
\email{gzhao@nao.cas.cn}

\correspondingauthor{Zhenxin Lei}
\email{zxlei@nao.cas.cn}

\author{Zhenxin Lei }
\affiliation{Key Laboratory of Optical Astronomy, National Astronomical Observatories, Chinese Academy of Sciences, Beijing 100012, China\\
}
\affiliation{College of Science, Shaoyang University, Shaoyang 422000, China\\}

\author{Jingkun Zhao }
\affiliation{Key Laboratory of Optical Astronomy, National Astronomical Observatories, Chinese Academy of Sciences, Beijing 100012, China\\
}

\author{P\'eter N\'emeth}
\affiliation{ Astronomical Institute of the Czech Academy of Sciences, CZ-251\,65, Ond\v{r}ejov, Czech Republic\\}
\affiliation{ Astroserver.org, 8533 Malomsok, Hungary\\
}

\author{Gang Zhao }
\affiliation{Key Laboratory of Optical Astronomy, National Astronomical Observatories, Chinese Academy of Sciences, Beijing 100012, China\\
}



\begin{abstract}
We selected 4593 hot subdwarf candidates from the Gaia DR2 
Hertzsprung-Russell (HR) diagram. By combining the sample with 
 LAMOST DR5,  we  identified 
294 hot subdwarf stars, including 169 sdB, 63 sdOB, 31 He-sdOB, 22 sdO, 7 He-sdO and 2 He-sdB stars.  
The atmospheric parameters (e.g., $T_\mathrm{eff}$, $\mathrm{log}\ g$, 
$\mathrm{log}(n\mathrm{He}/n\mathrm{H})$) are obtained by 
fitting the hydrogen (H) and helium (He) line profiles with 
synthetic spectra. Two distinct He sequences of 
hot subdwarf stars are clearly presented in  the $T_\mathrm{eff}$-
$\mathrm{log}\ g$ diagram. We found that the He-rich sequence 
consists of the bulk of sdB and sdOB stars as well as all of the He-sdB, He-sdO and He-sdOB stars 
in our samples, while all the stars in the He-weak sequence belong to the sdO 
spectral type,  combined with a few sdB and sdOB stars. 
We demonstrated that the combination of Gaia DR2 and LAMOST DR5 
allows one to uncover a huge number  of new hot subdwarf stars in our Galaxy.

\end{abstract}



\section{Introduction} \label{sec:intro}
Hot subdwarf stars consist of O and B type stars (spectral type sdO, sdB  and related objects) at a late stellar 
evolution stage (Heber 2009, Heber 2016). These stars have low stellar masses  
(e.g., about 0.5$\mathrm{M}_\odot$) and burn helium (He) in their cores or evolve 
off this evolution stage. Hot subdwarf stars present very high 
effective temperatures (e.g., $T_\mathrm{eff}\geq$ 20\,000 K) due to 
their  thin hydrogen (H) envelopes ($<0.01\mathrm{M}_\odot$, Heber 2016) or 
nearly pure He envelopes. Hot subdwarf stars play very important roles 
in the study of stellar evolution, especially for binary evolution. Furthermore, 
these stars are considered as the main sources of the ultraviolet (UV) upturn 
phenomena found in elliptical galaxies (O'Counnell 1999; Han et al. 2007; Jenkins 2013). 
Due to the diversity of H and He abundances in the atmosphere of hot subdwarf stars, 
they are ideal objects to test chemical diffusion theory as well (Michaud et al. 2008, 2011). 

Most sdB stars are considered to be formed in binary systems, since about half 
of these stars are found in close binaries (Maxted et al. 2001; Napiwotzki et al. 2004; 
Copperwheat et al. 2011). Employing a detailed binary population synthesis,  
Han et al. (2002, 2003) found that mass transfer through Roche lobe overflow (RLOF), 
common envelope (CE) ejection and merger of two He white dwarfs (WDs) can  
form sdB stars in binaries (also see Chen et al. 2013 and Xiong et al. 2017 for 
further study). Zhang et al. (2012, 2017) found that the 
merger of two He WDs or a He WD and a low-mass main-sequence (MS) star 
could produce single hot subdwarf stars, while Lei et al. (2015, 2016) demonstrated 
that tidally enhanced stellar wind in wide binaries is a viable formation 
channel for blue hook stars in massive globular clusters, which are the 
counterparts of field hot subdwarf stars. 

The number of known hot subdwarf stars remained at a low level until 
the Palomar-Green survey (PG, Green et al. 1986) of the northern Galactic 
hemisphere was released. With publicly available large surveys, 
such as Sloan Digital Sky survey (SDSS, York et al. 2000) and 
Large Sky Area Multi-Object Fibre Spectroscopic 
Telescope (LAMOST) survey (Cui et al. 2012; Zhao et al. 2006, 2012), etc, 
the number of known hot subdwarf stars increased largely in recent years.  Geier et al. (2017) 
compiled a catalogue of known hot subdwarf stars 
from the literature and unpublished databases, in which more than 
5000 hot subdwarf stars and candidates are listed. 

The method of candidate selection is a  very important 
step to identify hot subdwarf stars in large surveys. 
The conventional method  is to use 
color cuts followed by visual inspections. Employing this method, 
Geier et al. (2011) identified more than 1100 hot subdwarf stars 
from the SDSS survey; Vennes et al. (2011) and  N\'emeth et al. (2012) 
identified more than 200 hot subdwarf stars in Galaxy Evolution Explorer 
(GALEX) survey, while Luo et al. (2016)  found more than 100 
hot subdwarf stars in the LAMOST survey. In contrast to these studies, 
Bu et al. (2017) employed a  machine 
learning method to search for hot subdwarf stars in LAMOST DR1. 

Fortunately, the selection of hot subdwarf 
candidates has become much easier recently thanks to 
the second data release (DR2) of Gaia (Gaia collaboration et al. 2018a). 
The data of Gaia DR2  provide us the accurate positions, parallaxes and 
photometry for a large number of objects in our Galaxy. With these 
information, one can build a Hertzsprung-Russell (HR) diagram for a huge number 
of stars, in which nearly all stellar evolution sequences are presented clearly, including 
the hot subdwarf sequences. In this paper, we selected more than 4000 hot subdwarf candidates 
from the Gaia DR2 HR diagram built by Gaia collaboration et al. (2018b).  Combining with 
spectra from LAMOST DR5, we finally identified nearly 300 hot subdwarf stars from 
these candidates, among which 110  are newly 
identified in this study.  To confirm the identifications further, 
we obtained the atmospheric parameters of these hot subdwarf stars 
by fitting the profiles of H and He lines with synthetic spectra. 
The structure of this paper is as follows: In Section 2, 
we describe the hot subdwarf candidate selection procedure from 
the Gaia HR diagram and the spectral analysis. We 
give our results in Section 3. 
Finally, a discussion and summary are given in Section 4. 

\section{target selection and spectral analysis}
\subsection{Gaia DR2 database and HR diagram}
\begin{figure}
\centering

\begin{minipage}[c]{0.45\textwidth}
\includegraphics [width=70mm]{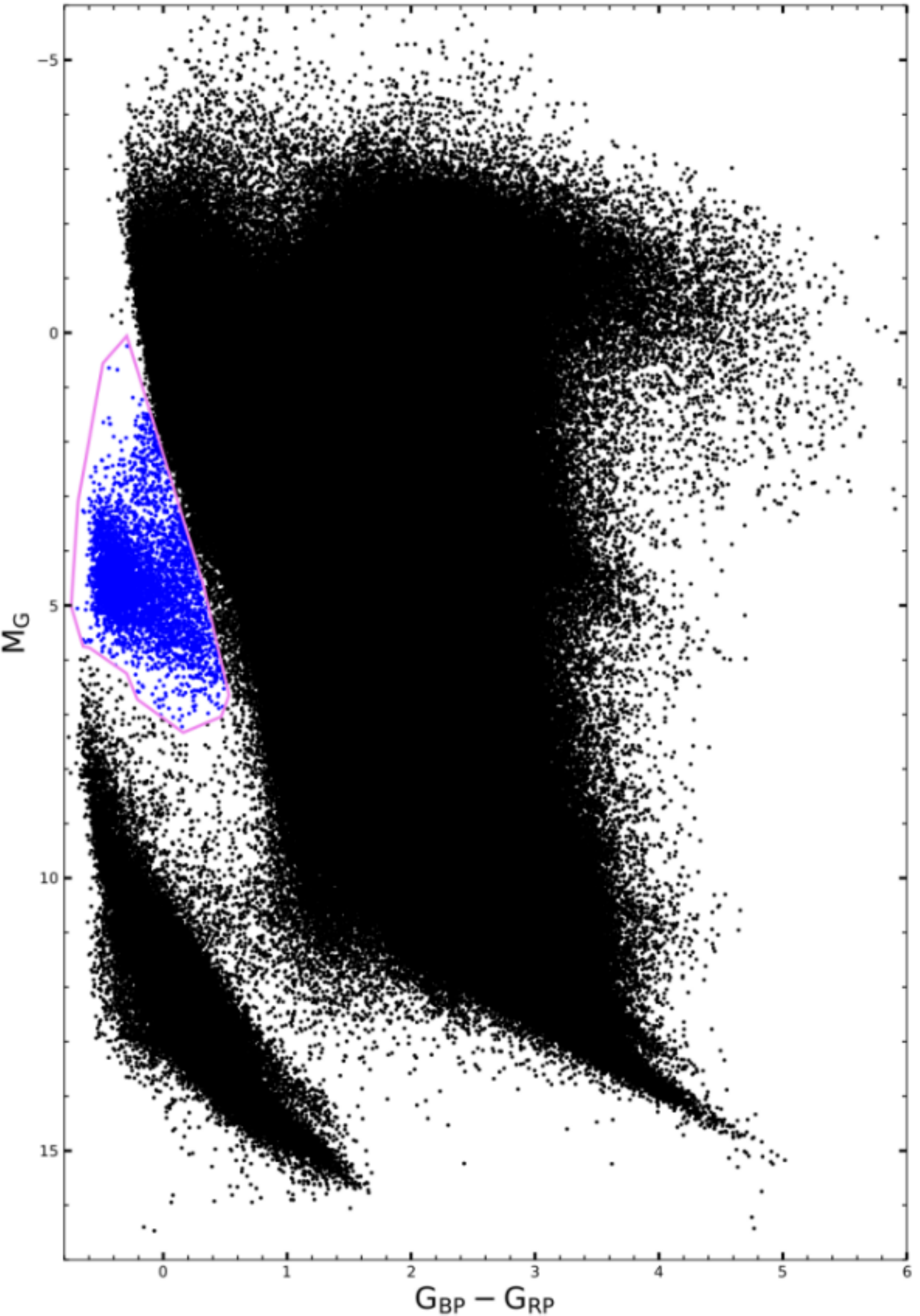}
\centerline{(a) }
\end{minipage}%
\begin{minipage}[c]{0.45\textwidth}
\includegraphics [width=70mm]{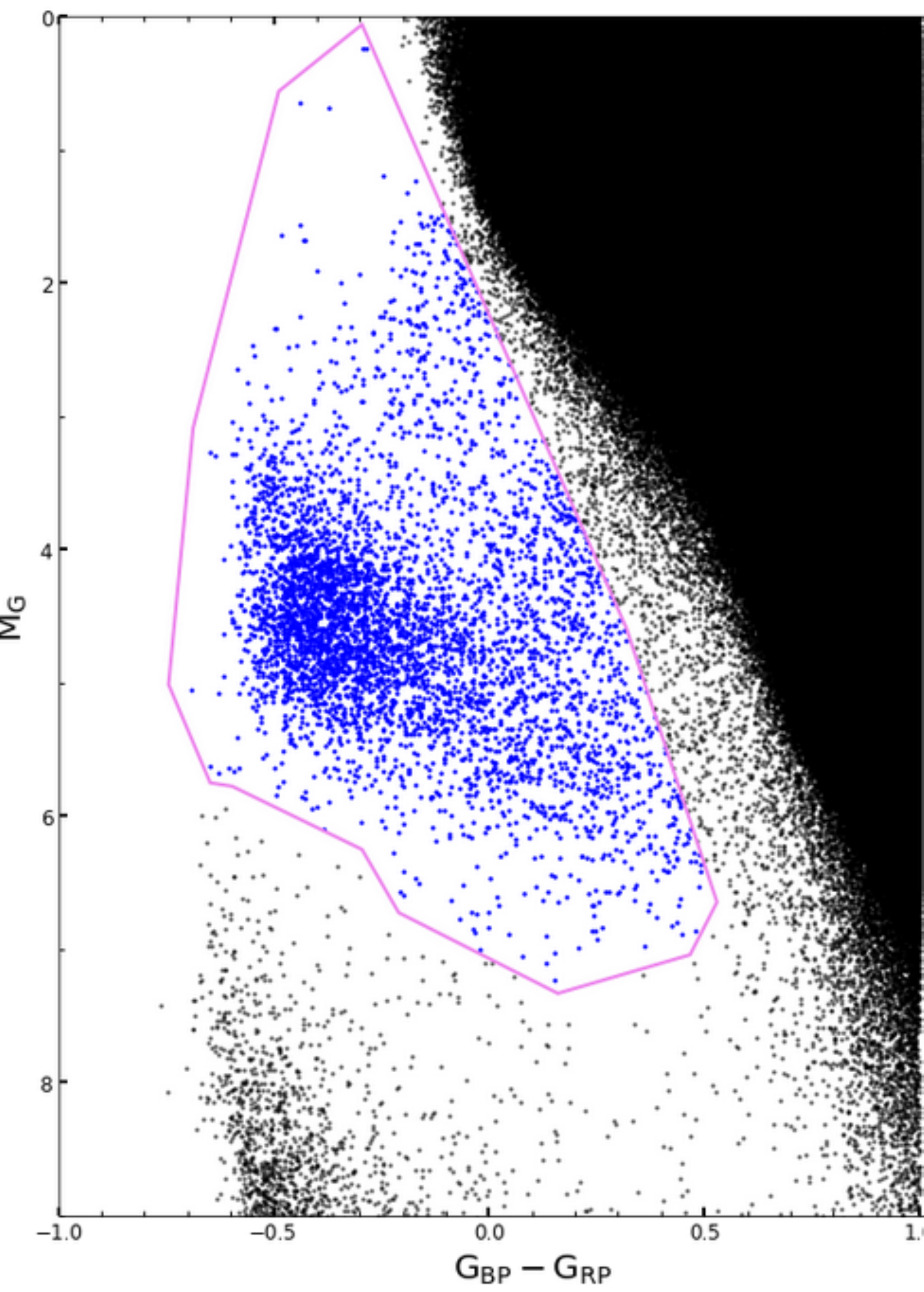}
\centerline{(b) }
\end{minipage}%

\caption{Panel (a): Gaia HR diagram from Gaia collaboration et al. (2018b). 4593 hot subdwarf 
candidates (blue dots) are selected visually within the boundaries 
of the hot subdwarf parameter space (pink polygon). Panel (b): Magnified drawing for  
the region of hot subdwarf candidates. }
\end{figure}

Gaia DR2 was released  on 25 April 2018 by the Gaia collaboration. Gaia DR2  consists 
of astrometric, photometric and radial velocity data, as well as variability and astrophysical parameters  for 
objects brighter than $21^{m}$ (Gaia Collaboration et al. 2018a). It contains approximately 1.7 billion sources with apparent magnitudes in white-light G-band (330-1050 nm), among which for about 1.3 billion sources 
are given  positions, parallaxes,  proper motions as well as photometry in the 
Blue (BP, 330-680 nm) and Red (RP, 630-1050 nm) band.  

With the accurate parallaxes and photometry provided by Gaia DR2, 
one can build a  HR diagram for a huge number of stars. 
Gaia collaboration et al. (2018b) built the Gaia HR diagram by selecting 
sources with the most precise parallaxes and photometry from Gaia DR2. 
To remove most of the artefacts (Arenou et al. 2018) while reserve plenty of genuine binaries 
between the white dwarf (WD) sequence and the main-sequence (MS)   
in the HR diagram, Arenou et al. (2018)  filtered out the astrometric 
excess noise by adopting the filter proposed in Appendix C of Lindergren et al. (2018). 
Gaia collaboration et al. (2018b)  estimated the absolute magnitude using 
$M_{G}=G+5+5\mathrm{log}_{10}(\varpi/1000)$, with $\varpi$ the 
parallax in milliarcseconds. To ensure the validity  of this equation, 
the relative uncertainty of the parallax is limited to be lower than 10\%. 
Furthermore, filters for the relative flux errors on the G, 
$G_\mathrm{BP}$ and $G_\mathrm{RP}$ magnitudes are also applied, which  
remove variable stars from the HR diagram. The detailed 
 selection filter  are presented 
 in Section 2 of Gaia collaboration et al. (2018b). 

Panel (a) of Fig 1 shows the Gaia HR diagram,  which 
contains 65\,921\,112 stars. Due to the effects of extinction, most of the 
evolution sequences, such as the MS, red giant branch (RGB) and asymptotic giant branch (AGB) are 
indistinguishable, only the WD  and hot subdwarf sequences show up clearly. 
However, one  can see that there are still many stars in the region 
between the main hot subdwarf sequence (e.g., structure near $\mathrm{M}_{G}\approx$ 5 
and $\mathrm{BP-RP} \approx$ -0.02 in Fig 1) and the wide MS. 
These objects could be real hot subdwarf stars, binaries or artefacts suffering from 
serious extinction issues (Arenou et al. 2018). To maximize the number of  hot subdwarf 
stars  in our study, we selected all the hot subdwarf candidates within  the  
pink polygon. Although this pink polygon is determined visually, it 
consists of the Extreme Horizontal Branch (EHB), which is the main hot subdwarf sequence, 
and nearly all the objects between the 
EHB and MS. There are in total 4593 hot subdwarf 
candidates in the polygon, which are marked by blue dots in Fig 1. 
Panel (b) of Fig 1 is the zoomed  region where our  
hot subdwarf candidates in the Gaia HR diagram were selected. 

\subsection{LAMOST DR5 }
LAMOST is a Chinese national scientific research facility operated 
by the National Astronomical Observatories,  
Chinese Academy of Sciences. It has a specially designed  reflecting Schmidt telescope 
with 4000 fibers in a field of view of 20 deg$^{2}$ in the sky. By July 2017, 
LAMOST has completed its pilot survey and completed its first five years of regular survey 
which was initiated in September 2012. 
After this six-year-survey,  LAMOST  obtained 9\,017\,844 spectra in total 
with a resolution ($\lambda/\Delta\lambda$) of 1800 in the 
wavelength range 3690-9100$\mathrm{\AA}$, which consist of 
8\,171\,443 stars, 153\,090 galaxies, 51\,133 quasars and 642\,178 unknown objects. 
These data make up the fifth data release (DR5) of LAMOST. 

We cross-matched our hot subdwarf candidates selected from the Gaia DR2 HR diagram 
with LAMOST DR5,  and found 734   
stars in common. To obtain reliable atmospheric parameters 
for these candidates, we removed the candidates with 
a  signal to noise ratio (SNR) less than 10 from our sample. Moreover, we  found 
that some candidates present obvious composite binary  
spectra, e.g., Mg I triplet lines at 5170 $\mathrm{\AA}$ 
and/or Ca II triplet lines at 8650 $\mathrm{\AA}$, and some of 
them present obvious H Balmer emission lines. These spectra 
have been removed from our sample as well.  
Such composite spectra will be analyzed in a forthcoming paper with appropriate models. 
After these procedures, we 
obtained 490 hot subdwarf candidates suitable for spectral analysis. 

\subsection{Fitting LAMOST spectra with synthetic spectra}
\begin{figure}
\centering 
\includegraphics [width=120mm]{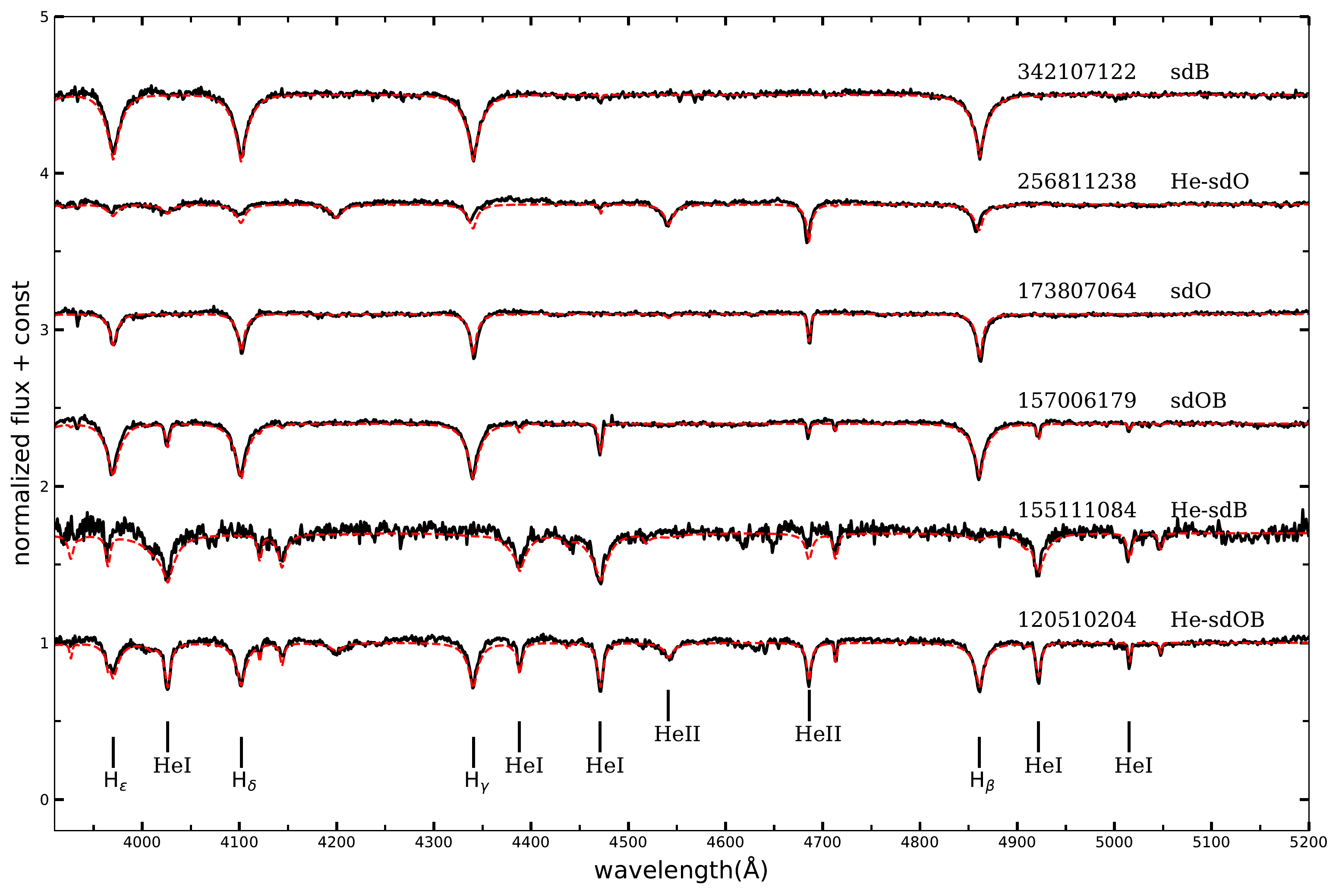} 
\caption{Normalized spectra for some hot subdwarf stars  identified 
in this study. Black solid curves are normalized observed spectra, and the 
red dashed curves are the best-fitting synthetic spectra. Some important 
H and He lines are marked by short vertical lines at the bottom of the figure. 
The LAMOST\_obsid and spectral classification type are presented on the right side 
for each spectra. }
\end{figure} 

Non-LTE model atmospheres with {\sc Tlusty} (version 204; Hubeny \& Lanz 2017) 
and synthetic spectra with {\sc Synspec} (version 49; Lanz et al. 2007)   
have been computed by N\'emeth et al. (2014), which 
were adopted in our study to fit the H and He profiles of our sample spectra. 
The spectral fitting was carried out by {\sc XTgrid}. 
This iterative procedure fits the observed spectrum by applying  successive adjustments 
to the synthetic spectrum 
following the gradient of steepest descent in the $\chi^2$ space ( see N\'emeth et al. 2012 for 
details). The synthetic spectra have been convolved with a 
Gaussian profile to match the resolution of the observed spectra,  and 
the parameter errors were estimated by mapping the $\Delta\chi^{2}$ field 
until the 60 per cent confidence level  at the given number of free parameters was reached. 

Fig 2 shows examples of fitting results in our study. In this figure, 
black solid curves are the normalized spectra for our candidates, while 
red dashed curves are the best-fitting synthetic spectra found by {\sc XTgrid}. 
Some important H and He lines are presented  by short vertical lines 
at the bottom of Fig 2 as well. Furthermore, the LAMOST\_obsid and spectral 
classification type for each star are also given in the right side of Fig 2. 
We followed the classification scheme of Geier et al. (2017, see their Table 1) 
to classify our hot subdwarf stars. 

The top spectrum in Fig 2 (i.e., 342107122) presents dominant H Balmer lines 
 but no He lines, which is a typical sdB star. The second 
spectrum from the top (i.e., 256811238) is a typical He-sdO star, because it shows 
dominant He II lines but very weak H Balmer and He I lines. 
The third spectrum from the top (i.e., 173807064) is classified  
as a sdO star, since it presents strong H balmer lines with weak He II line 
at 4686 $\mathrm{\AA}$, but no He I lines are detected. On the other hand, 
the fourth spectrum from the top (i.e., 157006179) is classified as  a 
sdOB star, because it presents strong H Balmer lines combined with both weak 
He I and He II lines. The fifth spectrum from the top (i.e., 155111084) shows dominant 
He I lines but no H Balmer and He II lines, which is typical to  He-sdB stars based on 
the classification scheme of Geier et al. (2017), while the bottom spectrum (i.e., 
120510204) shows dominant He I lines but also with He II and H Balmer lines, therefore 
we classified it as a He-sdOB star. 

\section{Results}
By fitting the observed spectra with synthetic spectra, we obtained the 
atmospheric parameters (e.g., $T_\mathrm{eff}$, $\mathrm{log}\ g$, 
$\mathrm{log}(n\mathrm{He}/n\mathrm{H})$) for the entire sample. 
Candidates with  $T_\mathrm{eff}<$ 20\,000 K  and  $\mathrm{log}\ g<$ 5.0 were  
considered as BHB stars, while candidates with $\mathrm{log}\ g<$ 4.5 were  
considered as B type MS stars (N\'emeth et al. 2012). 
By removing BHB, B type MS stars and duplicate objects, 
we identified 294 hot subdwarf stars, which include 169 sdB, 
63 sdOB, 31 He-sdOB, 22 sdO,   7 He-sdO and 2 He-sdB stars. 

All the parameters for our identified hot subdwarf stars are 
presented in Table 1. Columns 1-4 give the right ascension (RA), declination (Dec), LAMOST\_obsid and Gaia source\_id. Next, columns 5-8 give the $T_{\rm eff}$, $\log{g}$, $\log(n{\rm He}/n{\rm H})$ and radial velocity (RV) fitted by {\sc XTgrid}. Columns 9-11 list the apparent magnitude in the Gaia $G$ band, the SNR in the $u$ band and the spectral classification type, respectively.   
We  cross-matched  our results with the latest hot subdwarf catalogue of Geier et al. (2017), 
and got 184 common stars. We also cross-matched  our results with the hot subdwarf 
stars identified in N\'emeth et al. (2012) and Luo et al. (2016), and obtained 29 and 35  
common stars, respectively. The objects marked with $^{*}$ in 
Table 1 are common hot subdwarf stars  listed in 
the catalogue of Geier et al. (2017),  while objects labeled with  $^{\dagger}$  
and $^{\ddagger}$ are common  with  N\'emeth et al. (2012) 
and Luo et al. (2016) respectively. '$>$' in $\mathrm{log}(n\mathrm{He}/n\mathrm{H})$ 
denotes an upper limit of the He abundance, when  
{\sc XTgrid} could not find the error bars at the given quality of the spectra. 

\begin{table*}
\tiny

 \begin{minipage}{180mm}
  \caption{The information of 294 hot subdwarf stars identified in this study. From left to right, it gives 
  the right ascension (ra), declination (dec), LAMOST\_obsid,  and Gaia source\_id. Next the $T_{\rm eff}$, $\log{g}$, $\log(n{\rm He}/n{\rm H})$ and radial velocity (RV) are listed from the {\sc XTgrid} fits. Next, the apparent magnitudes in the Gaia $G$ band, the SNR for the $u$ band and the spectral classification types are listed, respectively.  }
  \end{minipage}\\
  \centering
    \begin{tabularx}{18.0cm}{lllcccccccccccccX}
\hline\noalign{\smallskip}
ra\tablenotemark{a}  & dec\tablenotemark{b}   &obsid\tablenotemark{c} & source\_id 
& $T_\mathrm{eff}$ &  $\mathrm{log}\ g$ & $\mathrm{log}(n\mathrm{He}/n\mathrm{H})$\tablenotemark{d}
& RV    &SNRU &$G$  & spclass\\
 LAMOST &  LAMOST& LAMOST&Gaia  &(K)&($\mathrm{cm\ s^{-2}}$)& & $\mathrm{km\ s^{-1}}$& &Gaia(mag) & \\
\hline\noalign{\smallskip}
0.0767$^{*}$ & 22.30079 & 385816105 & 2847977322031768832& 28810$\pm$280& 5.57$\pm$0.03 & -2.68$\pm$0.10 & 103.3 & 38.69 & 14.24 & sdB & \\
0.278039$^{*}$ & 11.010084 & 66605009$^{\ddagger}$ & 2765454164004531328& 27030$\pm$210& 5.48$\pm$0.02 & -2.70$\pm$0.06 & -35.6 & 21.58 & 13.61 & sdB & \\
0.981681$^{*}$ & 27.810404 & 492112073 & 2854091981071865344& 28000$\pm$90& 5.56$\pm$0.03 & -2.76$\pm$0.04 & 147.4 & 85.58 & 13.33 & sdB & \\
3.580483$^{*}$ & 22.40487 & 487408198 & 2800515322070491008& 36160$\pm$810& 5.35$\pm$0.07 & -3.12$\pm$0.34 & 7.8 & 12.21 & 13.43 & sdB & \\
8.943925$^{*}$ & 26.915094 & 157510209 & 2809421366255514112& 27150$\pm$520& 5.58$\pm$0.06 & -2.44$\pm$0.09 & 32.8 & 35.58 & 14.28 & sdB & \\
9.166037$^{*}$ & 37.931859$^{\dagger}$ & 75913167 & 368263110278860544& 37980$\pm$750& 6.30$\pm$0.08 & -3.00> & 213.2 & 15.23 & 14.14 & sdB & \\
9.507177 & 34.53228 & 14610118 & 365002268028633600& 40620$\pm$280& 5.55$\pm$0.07 & -0.27$\pm$0.07 & 73.6 & 15.63 & 13.90 & He-sdOB & \\
11.70128$^{*}$ & 19.93683 & 285308094 & 2801277176253968640& 32230$\pm$250& 5.89$\pm$0.06 & -2.27$\pm$0.06 & 18.6 & 39.38 & 14.98 & sdB & \\
11.918803$^{*}$ & 31.398766 & 186304225 & 360511037906831872& 30990$\pm$280& 5.75$\pm$0.04 & -2.24$\pm$0.05 & 81.0 & 45.85 & 14.81 & sdB & \\
12.321444$^{*}$ & 20.944565$^{\dagger}$ & 164803013 & 2801448085887897216& 27890$\pm$580& 5.34$\pm$0.07 & -2.44$\pm$0.20 & 6.2 & 20.10 & 14.56 & sdB & \\
12.455266$^{*}$ & 35.366938 & 14604029 & 363795760175504768& 35150$\pm$440& 5.78$\pm$0.06 & -1.54$\pm$0.06 & 11.9 & 29.80 & 14.82 & sdOB & \\
13.640353$^{*}$ & 37.970977 & 82010071 & 364742852004132736& 33680$\pm$530& 6.08$\pm$0.09 & -1.87$\pm$0.10 & 87.7 & 19.91 & 14.62 & sdOB & \\
14.602768$^{*}$ & 1.909744 & 8206211$^{\ddagger}$ & 2537690089791382656& 33290$\pm$600& 5.53$\pm$0.08 & -1.90$\pm$0.10 & -53.7 & 22.52 & 15.13 & sdOB & \\
15.305503$^{*}$ & 31.432091 & 96302063 & 312628749626419328& 27150$\pm$250& 5.48$\pm$0.02 & -2.89$\pm$0.10 & -41.9 & 50.74 & 13.08 & sdB & \\
16.090314$^{*}$ & 4.226959 & 55112148$^{\ddagger}$ & 2551900379931546240& 27760$\pm$180& 5.55$\pm$0.02 & -2.72$\pm$0.05 & 30.3 & 31.07 & 12.05 & sdB & \\
16.203409$^{*}$ & 36.461784 & 82001006 & 369576820516013824& 32270$\pm$70& 5.77$\pm$0.02 & -1.64$\pm$0.02 & 10.7 & 113.49 & 12.40 & sdOB & \\
17.44054$^{*}$ & 37.760636 & 23008132 & 369573831218501376& 29070$\pm$160& 5.49$\pm$0.03 & -3.44$\pm$0.43 & 33.8 & 13.43 & 13.87 & sdB & \\
19.870333$^{*}$ & 49.019264$^{\dagger}$ & 82712038$^{\ddagger}$ & 400211958948711552& 43170$\pm$520& 5.70$\pm$0.11 & 0.07$\pm$0.09 & -1.4 & 11.41 & 13.37 & He-sdOB & \\
21.65453$^{*}$ & 16.1365 & 363615074 & 2592234628262141312& 30490$\pm$130& 5.72$\pm$0.01 & -2.42$\pm$0.03 & 33.8 & 96.43 & 14.31 & sdB & \\
23.849927$^{*}$ & 37.34115 & 401302195 & 322575344128842624& 55470$\pm$950& 5.33$\pm$0.26 & 0.55$\pm$0.13 & 26.6 & 98.86 & 14.11 & He-sdO & \\
24.109146$^{*}$ & 11.658952 & 482304144 & 2585744516065582848& 29590$\pm$510& 5.71$\pm$0.02 & -2.26$\pm$0.05 & 43.9 & 32.60 & 12.30 & sdB & \\
27.72618 & 31.129651 & 14713190 & 304311150320629376& 28520$\pm$270& 5.74$\pm$0.03 & -1.73$\pm$0.05 & 76.3 & 17.08 & 14.32 & sdB & \\
28.324146 & 35.701356 & 354713006 & 330261789400007040& 29860$\pm$430& 6.06$\pm$0.08 & -2.67$\pm$0.23 & -39.9 & 10.37 & 15.73 & sdB & \\
28.591037$^{*}$ & 49.089103$^{\dagger}$ & 503213129 & 357580873780641920& 33850$\pm$100& 5.70$\pm$0.04 & -2.00$\pm$0.07 & 83.8 & 12.42 & 13.47 & sdB & \\
29.647084 & 55.084686 & 380703127 & 504485110543295616& 51300$\pm$3500& 5.24$\pm$0.04 & -2.31$\pm$0.12 & 21.1 & 47.42 & 15.52 & sdO & \\
30.5077 & 56.7284 & 15302097 & 505063728541561344& 28920$\pm$590& 5.62$\pm$0.07 & -2.47$\pm$0.11 & -39.2 & 12.48 & 15.24 & sdB & \\
31.537384$^{*}$ & 37.647643 & 510514015 & 330722897089067392& 27630$\pm$220& 5.41$\pm$0.03 & -2.85$\pm$0.12 & 32.0 & 28.37 & 14.17 & sdB & \\
33.110821$^{*}$ & 1.921483 & 290113234 & 2514278566658235776& 31650$\pm$370& 5.63$\pm$0.02 & -2.71$\pm$0.12 & -41.1 & 30.45 & 14.19 & sdB & \\
33.708238$^{*}$ & 23.331535 & 379610233 & 101096658200679936& 26600$\pm$240& 5.63$\pm$0.04 & -3.16$\pm$0.23 & 34.6 & 35.94 & 15.67 & sdB & \\
34.42565 & 28.058199 & 173213178 & 107665724780053376& 33120$\pm$110& 5.92$\pm$0.04 & -1.61$\pm$0.03 & 10.1 & 86.75 & 11.78 & sdOB & \\
38.216289$^{*}$ & 44.190928$^{\dagger}$ & 180412223 & 341195058149873024& 34400$\pm$570& 5.75$\pm$0.06 & -1.71$\pm$0.04 & -60.2 & 36.48 & 14.27 & sdOB & \\
38.317737 & 8.380032 & 364610005 & 20063858119453824& 29830$\pm$110& 5.43$\pm$0.03 & -2.39$\pm$0.07 & 31.2 & 17.22 & 15.38 & sdB & \\
40.844775$^{*}$ & 4.843528 & 293409213 & 6052403489630720& 34790$\pm$110& 6.02$\pm$0.03 & 0.09$\pm$0.03 & -64.8 & 103.08 & 14.09 & He-sdOB & \\
41.89582$^{*}$ & 36.763992 & 82312150$^{\ddagger}$ & 141428363911552384& 43520$\pm$210& 5.89$\pm$0.13 & 0.52$\pm$0.03 & 74.4 & 48.24 & 12.98 & He-sdOB & \\
45.315923$^{*}$ & -2.670405 & 367406216 & 5186434732342771968& 26820$\pm$230& 5.39$\pm$0.02 & -2.46$\pm$0.05 & 18.6 & 66.16 & 13.84 & sdB & \\
49.487181 & 32.497341 & 76704053 & 125093503772825856& 34250$\pm$230& 5.97$\pm$0.05 & -1.62$\pm$0.05 & 74.7 & 15.51 & 15.72 & sdOB & \\
50.662367$^{*}$ & 46.076037 & 408714065 & 242734753958329728& 41490$\pm$50& 5.76$\pm$0.06 & 1.19$\pm$0.09 & 69.8 & 15.61 & 13.66 & He-sdOB & \\
50.989343 & 42.270805 & 353410213 & 240978868246214400& 35350$\pm$440& 5.81$\pm$0.08 & -1.45$\pm$0.07 & 14.8 & 17.34 & 15.77 & sdOB & \\
51.907159$^{*}$ & 38.135982 & 265511138 & 234861632224343424& 26010$\pm$150& 5.52$\pm$0.01 & -2.84$\pm$0.05 & -104.5 & 62.88 & 14.22 & sdB & \\
53.288922 & 34.208582 & 292911128 & 221005208733887488& 34540$\pm$420& 5.64$\pm$0.05 & -1.67$\pm$0.04 & 26.4 & 21.33 & 15.79 & sdOB & \\
54.589981 & 41.734607 & 296711237 & 237331822538478592& 27200$\pm$140& 5.59$\pm$0.03 & -1.51$\pm$0.03 & 9.1 & 30.37 & 15.21 & sdB & \\
54.696112 & 41.573393 & 296715043 & 237133566847520000& 22240$\pm$320& 5.76$\pm$0.03 & -3.00> & 100.8 & 30.84 & 15.11 & sdB & \\
54.944312 & 35.546154 & 265506073 & 221375331836119552& 64650$\pm$9600& 5.61$\pm$0.05 & -1.58$\pm$0.08 & 23.0 & 26.89 & 15.84 & sdO & \\
56.022214 & 22.072953 & 249811209 & 64368713521485568& 28130$\pm$440& 5.59$\pm$0.06 & -2.74$\pm$0.09 & 29.7 & 32.31 & 14.40 & sdB & \\
57.933712 & 36.871319 & 181511115 & 220091342774317056& 28170$\pm$320& 5.37$\pm$0.08 & -2.58$\pm$0.43 & 31.8 & 10.59 & 16.18 & sdB & \\
58.380531$^{*}$ & 10.751119 & 159408059 & 3303143628052101120& 42400$\pm$230& 5.70$\pm$0.05 & -0.01$\pm$0.06 & -30.5 & 62.97 & 12.83 & He-sdOB & \\
58.809479$^{*}$ & 10.470014 & 159306197 & 3302955130527310464& 26340$\pm$150& 5.33$\pm$0.01 & -2.68$\pm$0.06 & 35.0 & 51.46 & 14.15 & sdB & \\
59.276419 & 36.474847 & 397401216 & 220172981513287936& 84550$\pm$3640& 5.58$\pm$0.04 & -1.30$\pm$0.15 & 25.6 & 22.98 & 15.49 & sdO & \\
60.052436 & 36.596069 & 397401025 & 219511728348315520& 22480$\pm$120& 5.36$\pm$0.02 & -3.07$\pm$0.22 & 18.3 & 20.86 & 14.89 & sdB & \\
60.197058 & 17.311691 & 280513233 & 46608061680943872& 28750$\pm$290& 5.57$\pm$0.04 & -3.15$\pm$0.20 & 23.7 & 17.22 & 15.45 & sdB & \\
60.675411 & 31.397095 & 182113038 & 169258102724229888& 36410$\pm$340& 6.00$\pm$0.10 & -1.52$\pm$0.08 & -48.6 & 13.70 & 15.83 & sdOB & \\
61.093981 & 37.408948 & 397507109 & 225573267233289600& 34470$\pm$290& 5.72$\pm$0.10 & -1.43$\pm$0.09 & 8.5 & 19.94 & 16.20 & sdOB & \\
63.958993$^{*}$ & 1.905817 & 174708071 & 3259060049366022400& 32920$\pm$90& 5.78$\pm$0.02 & -1.71$\pm$0.03 & 9.8 & 87.29 & 14.04 & sdOB & \\
64.820704 & 36.473358 & 162001168 & 177389712764641920& 42340$\pm$250& 5.47$\pm$0.05 & -0.38$\pm$0.04 & -7.8 & 39.03 & 14.62 & He-sdOB & \\
65.998459 & 34.463668 & 204704027 & 176209799348981376& 35580$\pm$220& 5.79$\pm$0.04 & -1.49$\pm$0.04 & 82.6 & 41.01 & 15.19 & sdOB & \\
66.040252 & 8.554802 & 425604162 & 3299455110137611776& 27210$\pm$190& 5.27$\pm$0.01 & -1.32$\pm$0.02 & -0.5 & 142.89 & 12.53 & sdB & \\
67.389212$^{*}$ & 7.697824 & 527312249 & 3287144634355536768& 44080$\pm$1010& 5.62$\pm$0.04 & -2.40$\pm$0.08 & 18.2 & 34.29 & 15.07 & sdO & \\
67.80107$^{*}$ & 42.986157 & 420301136 & 252329740960393984& 26690$\pm$180& 5.63$\pm$0.01 & -2.71$\pm$0.06 & 32.8 & 63.34 & 14.56 & sdB & \\
71.237089$^{*}$ & 14.363909 & 282807077 & 3308791681845675136& 32660$\pm$320& 5.66$\pm$0.05 & -2.49$\pm$0.13 & -54.0 & 16.33 & 14.97 & sdB & \\
71.814516$^{*}$ & 20.666174 & 283416013 & 3411847601045302144& 30250$\pm$700& 5.51$\pm$0.12 & -1.88$\pm$0.24 & 29.2 & 13.79 & 15.55 & sdB & \\
72.196713 & -4.00472 & 392308023 & 3200857538788110080& 29370$\pm$50& 5.65$\pm$0.01 & -1.97$\pm$0.03 & 17.9 & 44.84 & 13.36 & sdB & \\
73.990376 & 46.318864 & 364911218 & 254631847728048384& 27250$\pm$430& 5.45$\pm$0.04 & -2.73$\pm$0.23 & -108.9 & 14.90 & 15.72 & sdB & \\
75.660378 & 45.720627 & 364912013 & 206657161813612288& 32840$\pm$210& 6.04$\pm$0.05 & -1.92$\pm$0.07 & -52.1 & 13.61 & 15.67 & sdB & \\
76.68493 & 11.604413 & 407202008 & 3388118112893920000& 32480$\pm$210& 5.34$\pm$0.12 & -1.31$\pm$0.06 & 72.3 & 26.66 & 15.42 & sdOB & \\
77.306642 & 33.503376 & 209916235 & 181774221540061824& 25640$\pm$200& 5.32$\pm$0.01 & -2.13$\pm$0.05 & -59.6 & 21.09 & 15.44 & sdB & \\
77.51154$^{*}$ & 12.879591 & 407205159 & 3388371103647349632& 32300$\pm$160& 6.06$\pm$0.05 & -2.09$\pm$0.07 & 7.8 & 12.46 & 16.07 & sdB & \\
77.963858$^{*}$ & 41.691511 & 252814174 & 201146826149598464& 31530$\pm$350& 6.26$\pm$0.15 & -2.56$\pm$0.23 & -21.3 & 12.06 & 15.58 & sdB & \\
78.457499$^{*}$ & 17.700698 & 317806011 & 3395250473024213120& 30990$\pm$200& 5.50$\pm$0.05 & -1.97$\pm$0.04 & 11.0 & 11.26 & 14.06 & sdB & \\
80.378419 & 37.362848 & 162512222 & 184468780942344960& 26600$\pm$300& 5.72$\pm$0.05 & -3.00> & 108.7 & 11.86 & 14.28 & sdB & \\
80.96406 & 14.078763 & 374402025 & 3390235531771243904& 21770$\pm$120& 5.01$\pm$0.05 & -2.17$\pm$0.06 & 31.4 & 30.53 & 15.86 & sdB & \\
\hline\noalign{\smallskip} 
  \end{tabularx}
 \tablenotetext{a}{Stars labeled with $\ast$ also appear in the hot subdwarf catalogue of Geier et al. (2017).} \tablenotetext{b}{Stars labeled with $\dagger$ also appear in N\'emeth et al. (2012).} 
 \tablenotetext{c}{Stars labeled with $\ddagger$ also appear in Luo et al. (2016).}
 \tablenotetext{d}{"$>$" denotes a upper limit of $\mathrm{log}(n\mathrm{He}/n\mathrm{H})$ for the object.} 

\end{table*}

\setcounter{table}{0}
\begin{table*}
\tiny

 \begin{minipage}{180mm}
  \caption{ continued. }
  \end{minipage}\\
  \centering
    \begin{tabularx}{18.0cm}{lllcccccccccccccX}
\hline\noalign{\smallskip}
ra\tablenotemark{a}  & dec\tablenotemark{b}   &obsid\tablenotemark{c} & source\_id 
& $T_\mathrm{eff}$ &  $\mathrm{log}\ g$ & $\mathrm{log}(n\mathrm{He}/n\mathrm{H})$\tablenotemark{d}
& RV    &SNRU &$G$ & spclass\\
 LAMOST &  LAMOST& LAMOST&Gaia  &(K)&($\mathrm{cm\ s^{-2}}$)& & $\mathrm{km\ s^{-1}}$& &Gaia(mag) & \\
\hline\noalign{\smallskip}
83.279828 & 7.506107 & 208913243 & 3334223901191558784& 35370$\pm$1050& 5.89$\pm$0.06 & -1.66$\pm$0.07 & -57.1 & 33.66 & 13.81 & sdOB & \\
83.338169 & 16.113207 & 258002165 & 3396876131029433472& 33930$\pm$290& 5.92$\pm$0.05 & -1.66$\pm$0.05 & 80.1 & 28.37 & 15.23 & sdB & \\
83.465224 & 19.241956 & 279007127 & 3400946935393688576& 27170$\pm$250& 5.36$\pm$0.03 & -3.03$\pm$0.19 & -34.2 & 17.48 & 15.76 & sdB & \\
84.235335$^{*}$ & 39.92188 & 42704127 & 190970433715259904& 37890$\pm$330& 5.57$\pm$0.06 & -0.63$\pm$0.06 & -0.5 & 17.17 & 13.92 & He-sdOB & \\
85.740707 & 39.197564 & 284213200 & 190164354254168192& 36580$\pm$400& 5.68$\pm$0.06 & -1.49$\pm$0.06 & 17.6 & 11.36 & 14.94 & sdOB & \\
86.189735 & 30.648213 & 408809093 & 3444598749605672064& 28520$\pm$150& 5.59$\pm$0.02 & -2.50$\pm$0.04 & 33.9 & 41.89 & 12.43 & sdB & \\
86.235913 & 35.826651 & 319909136 & 3455626713794023680& 45570$\pm$180& 5.60$\pm$0.04 & 0.47$\pm$0.05 & -0.4 & 37.80 & 15.26 & He-sdOB & \\
86.775955 & 40.139606 & 321102181 & 191727718051389312& 32880$\pm$300& 5.88$\pm$0.05 & -1.24$\pm$0.05 & 6.6 & 21.52 & 15.76 & sdOB & \\
87.96384 & 22.076954 & 107916117 & 3424202297812639360& 29210$\pm$190& 5.60$\pm$0.02 & -2.22$\pm$0.04 & 13.7 & 24.41 & 13.17 & sdB & \\
88.167219$^{*}$ & 19.820386 & 195804131 & 3398928751737464192& 80890$\pm$14680& 5.58$\pm$0.11 & -2.21$\pm$0.28 & -29.2 & 28.36 & 14.86 & sdO & \\
88.302057$^{*}$ & 19.397253 & 195805160 & 3398852881140229632& 29140$\pm$190& 5.56$\pm$0.02 & -2.80$\pm$0.16 & -40.2 & 38.34 & 14.53 & sdB & \\
88.453581 & 32.93382 & 66503084 & 3451217092350738944& 30500$\pm$80& 5.66$\pm$0.02 & -2.18$\pm$0.03 & -51.8 & 50.97 & 14.17 & sdB & \\
89.391565 & 14.204803 & 271101097 & 3346893264443563776& 37240$\pm$290& 5.59$\pm$0.03 & -1.43$\pm$0.04 & -52.9 & 33.36 & 14.55 & sdOB & \\
89.707524 & 18.519722 & 195807009 & 3350560685476251648& 29880$\pm$370& 5.40$\pm$0.08 & -2.60$\pm$0.22 & -35.5 & 19.27 & 15.65 & sdB & \\
90.129088 & 29.148612 & 212009005 & 3431631727942970240& 35000$\pm$140& 5.34$\pm$0.03 & -0.87$\pm$0.03 & -64.1 & 39.35 & 13.19 & He-sdOB & \\
90.157054 & 36.799292 & 381508201 & 3456632255835216640& 26300$\pm$1160& 5.30$\pm$0.14 & -3.24$\pm$0.29 & -33.8 & 13.38 & 15.46 & sdB & \\
90.257966 & 42.511474 & 321109060 & 960904069540367232& 35770$\pm$270& 5.90$\pm$0.05 & -1.51$\pm$0.04 & 14.8 & 30.16 & 15.28 & sdOB & \\
91.22292 & 9.467549 & 526403092 & 3329611552994806272& 47960$\pm$640& 5.70$\pm$0.11 & 0.17$\pm$0.05 & -104.4 & 20.84 & 14.56 & He-sdO & \\
92.665509 & 7.903821 & 526401140 & 3328247780618525568& 27850$\pm$370& 5.47$\pm$0.03 & -2.76$\pm$0.11 & -40.0 & 25.65 & 15.06 & sdB & \\
92.742942 & 13.375245 & 400104005 & 3344349578650746880& 29070$\pm$320& 5.61$\pm$0.09 & -3.00> & -16.9 & 18.48 & 16.02 & sdB & \\
92.845855 & 55.870869 & 497610101 & 997207297787944064& 29780$\pm$470& 5.46$\pm$0.07 & -2.46$\pm$0.13 & 9.5 & 25.00 & 15.81 & sdB & \\
93.304375$^{*}$ & 34.933493 & 187212172 & 3452480293768439296& 28970$\pm$280& 5.49$\pm$0.07 & -3.00> & -35.5 & 27.35 & 14.47 & sdB & \\
93.636538 & 11.061654 & 398805096 & 3330105641733251840& 31300$\pm$220& 5.47$\pm$0.04 & -1.83$\pm$0.06 & 12.0 & 13.18 & 16.41 & sdOB & \\
94.475909 & 23.631485 & 176413170 & 3425561981379261184& 36120$\pm$470& 5.70$\pm$0.07 & -1.52$\pm$0.02 & 7.5 & 18.31 & 15.39 & sdOB & \\
94.486139 & 18.83019 & 555306021 & 3373814561832902272& 26600$\pm$480& 5.47$\pm$0.05 & -2.91$\pm$0.15 & -39.9 & 24.90 & 13.60 & sdB & \\
94.678473$^{*}$ & 21.592955 & 483207093 & 3376639486380091008& 31380$\pm$60& 5.80$\pm$0.02 & -2.95$\pm$0.08 & -35.2 & 59.50 & 14.67 & sdB & \\
94.864687$^{*}$ & 35.382661 & 220504054 & 3452740362628172160& 35590$\pm$210& 5.53$\pm$0.02 & -3.00> & -58.4 & 55.90 & 15.05 & sdOB & \\
95.205672 & 19.308045 & 256512020 & 3373940219694568576& 30420$\pm$100& 5.71$\pm$0.02 & -2.05$\pm$0.07 & -54.4 & 16.85 & 14.96 & sdB & \\
96.592792 & 4.073252 & 442708048 & 3131204469305148288& 26620$\pm$70& 5.53$\pm$0.01 & -2.78$\pm$0.05 & 23.6 & 68.24 & 12.32 & sdB & \\
96.770481$^{*}$ & 34.969325 & 74716039 & 942463434178440576& 25330$\pm$540& 5.30$\pm$0.05 & -3.34$\pm$0.30 & 29.6 & 12.20 & 14.43 & sdB & \\
97.152155 & 32.842084 & 74703090 & 3439296187280018432& 43240$\pm$610& 5.49$\pm$0.09 & 0.11$\pm$0.07 & -5.9 & 23.47 & 14.71 & He-sdOB & \\
97.640728 & 21.324502 & 274309125 & 3376057123175008256& 21460$\pm$240& 5.32$\pm$0.02 & -2.36$\pm$0.20 & -38.6 & 15.27 & 15.33 & sdB & \\
98.043207 & 28.178276 & 75709065 & 3434099173772378112& 45110$\pm$610& 5.44$\pm$0.08 & 0.40$\pm$0.07 & -0.7 & 18.97 & 15.10 & He-sdOB & \\
98.497308 & 32.553966 & 74704076 & 3436238720320660224& 26010$\pm$1210& 5.86$\pm$0.11 & -2.97$\pm$0.42 & -57.6 & 13.01 & 15.16 & sdB & \\
99.967315 & 51.950267 & 117705238 & 992534888766785024& 26230$\pm$240& 5.97$\pm$0.01 & -2.99$\pm$0.13 & -103.2 & 40.08 & 11.96 & sdB & \\
100.45235$^{*}$ & 42.805621 & 422811236 & 963881581386403072& 28140$\pm$320& 5.56$\pm$0.06 & -2.77$\pm$0.14 & -100.8 & 21.89 & 14.20 & sdB & \\
100.81184 & 13.965445 & 415715194 & 3352890447538648064& 32560$\pm$260& 5.76$\pm$0.04 & -2.18$\pm$0.05 & -51.5 & 24.60 & 14.90 & sdB & \\
100.88299$^{*}$ & 14.157207 & 415715176 & 3355897092143865088& 35930$\pm$180& 5.57$\pm$0.02 & -3.00> & -54.3 & 27.31 & 14.83 & sdB & \\
100.948932$^{*}$ & 32.029482 & 438514106 & 937248901501446272& 26880$\pm$70& 5.21$\pm$0.01 & -2.70$\pm$0.05 & -37.5 & 29.84 & 13.83 & sdB & \\
101.57652$^{*}$ & 29.337016 & 2604051 & 3386622158606040960& 37930$\pm$360& 6.00$\pm$0.05 & -3.00> & -101.5 & 74.81 & 13.59 & sdB & \\
103.09651 & 9.965476 & 392002223 & 3158419684195782656& 70180$\pm$1800& 5.78$\pm$0.04 & -1.44$\pm$0.10 & -44.4 & 88.90 & 14.22 & sdO & \\
103.18157 & 40.293805 & 296115026 & 950627174872167808& 68730$\pm$1180& 5.21$\pm$0.03 & -0.29$\pm$0.05 & 96.3 & 92.62 & 14.13 & He-sdO & \\
103.351374 & 33.059526 & 438511176 & 937565831431448192& 33990$\pm$100& 5.49$\pm$0.01 & -1.49$\pm$0.03 & -268.2 & 107.46 & 12.11 & sdOB & \\
103.69431 & 24.82412 & 81505213 & 3381286602335612416& 61260$\pm$2250& 5.10$\pm$0.02 & -2.00$\pm$0.07 & -40.6 & 55.28 & 13.99 & sdO & \\
103.88743 & 22.063784 & 95406046 & 3378454092881858816& 47580$\pm$940& 5.69$\pm$0.06 & -1.70$\pm$0.08 & -50.3 & 32.27 & 13.70 & sdOB & \\
104.56965 & 9.728642 & 29001011 & 3157967411257049728& 36690$\pm$1110& 5.01$\pm$0.07 & -1.74$\pm$0.05 & -53.1 & 18.19 & 13.59 & sdOB & \\
106.31091 & 22.355964 & 174911065 & 3367827072481260416& 28550$\pm$280& 5.62$\pm$0.04 & -2.71$\pm$0.12 & -36.8 & 28.70 & 15.12 & sdB & \\
106.82751 & 14.016182 & 372807096 & 3161497015379649280& 39820$\pm$1010& 5.61$\pm$0.07 & -1.95$\pm$0.14 & 8.2 & 32.01 & 15.42 & sdOB & \\
108.01 & 11.559014 & 128701029 & 3159937564294110080& 25790$\pm$160& 5.43$\pm$0.01 & -2.65$\pm$0.04 & 28.7 & 48.56 & 12.46 & sdB & \\
108.62448$^{*}$ & 22.283159 & 177904013 & 3367249485278272000& 41840$\pm$10& 6.10$\pm$0.06 & 0.76$\pm$0.04 & -19.3 & 257.09 & 11.61 & He-sdOB & \\
109.740975$^{*}$ & 2.887462 & 492402100 & 3139042514040855296& 26300$\pm$160& 5.23$\pm$0.02 & -2.46$\pm$0.10 & -34.5 & 16.45 & 14.89 & sdB & \\
110.666731 & 9.231818 & 369611044 & 3155626997678101248& 48010$\pm$2010& 5.79$\pm$0.09 & -2.71$\pm$0.29 & 18.7 & 22.22 & 13.05 & sdO & \\
110.96446$^{*}$ & 30.321259 & 214908131$^{\ddagger}$ & 885963418573729792& 32400$\pm$190& 5.91$\pm$0.04 & -1.53$\pm$0.04 & 17.5 & 44.35 & 15.04 & sdOB & \\
111.22127$^{*}$ & 19.391139 & 270103045 & 3362730178953977472& 26520$\pm$120& 5.53$\pm$0.02 & -2.54$\pm$0.05 & -56.0 & 53.35 & 14.67 & sdB & \\
111.2289 & 15.280799 & 397603157 & 3167124522048560128& 32490$\pm$400& 5.88$\pm$0.08 & -1.77$\pm$0.08 & -63.2 & 39.25 & 15.54 & sdOB & \\
111.65522 & 8.558556 & 446301096 & 3143342600936696960& 29920$\pm$130& 5.64$\pm$0.02 & -3.00> & -158.5 & 50.05 & 15.03 & sdB & \\
113.35669$^{*}$ & 36.882668 & 331004092 & 898625493262383104& 33110$\pm$80& 5.94$\pm$0.02 & -2.42$\pm$0.04 & -39.4 & 58.95 & 15.10 & sdB & \\
114.182985 & -3.195907 & 436206051 & 3060798513530161920& 28890$\pm$400& 5.54$\pm$0.04 & -1.11$\pm$0.04 & -65.6 & 32.14 & 13.75 & sdB & \\
114.48439 & 31.279597 & 91905056 & 880252005422941440& 31000$\pm$290& 5.49$\pm$0.05 & -2.42$\pm$0.09 & 12.0 & 10.95 & 13.58 & sdB & \\
114.51894$^{*}$ & 26.412458$^{\dagger}$ & 73906248 & 869050009160765056& 27730$\pm$660& 5.55$\pm$0.07 & -3.14$\pm$0.65 & -99.8 & 10.14 & 14.58 & sdB & \\
116.14643$^{*}$ & 30.352421$^{\dagger}$ & 120407017 & 879166065892619904& 29100$\pm$350& 5.45$\pm$0.06 & -2.86$\pm$0.10 & 32.5 & 36.33 & 14.74 & sdB & \\
116.28351$^{*}$ & 38.18521 & 279210190 & 919963474904219776& 29310$\pm$270& 5.30$\pm$0.06 & -0.93$\pm$0.04 & 1.0 & 40.58 & 14.84 & sdB & \\
116.46916 & 25.85761 & 183702062 & 868817943488022016& 26210$\pm$250& 5.47$\pm$0.03 & -2.31$\pm$0.09 & 34.7 & 46.95 & 14.75 & sdB & \\
117.23262$^{*}$ & 30.713059$^{\dagger}$ & 120407139 & 879237740307303040& 31080$\pm$90& 5.65$\pm$0.02 & -1.98$\pm$0.04 & -122.1 & 42.67 & 14.06 & sdB & \\
117.91362 & 6.768001 & 120510204 & 3144058245567712512& 39600$\pm$80& 5.61$\pm$0.07 & -0.14$\pm$0.02 & 4.4 & 47.16 & 13.50 & He-sdOB & \\
118.55157$^{*}$ & 29.832504 & 2711238 & 877431345781684480& 31270$\pm$1010& 5.82$\pm$0.19 & -1.75$\pm$0.16 & -42.3 & 15.02 & 14.57 & sdB & \\
119.01571$^{*}$ & 22.441942$^{\dagger}$ & 173807064 & 674255203425267584& 65800$\pm$1420& 5.48$\pm$0.01 & -2.00$\pm$0.10 & -40.1 & 79.20 & 14.19 & sdO & \\
119.53142$^{*}$ & -4.534807$^{\dagger}$ & 170213155 & 3068507812327598208& 41370$\pm$60& 5.87$\pm$0.07 & 0.36$\pm$0.05 & -98.2 & 54.63 & 13.06 & He-sdOB & \\
121.73617$^{*}$ & 15.455543$^{\dagger}$ & 392709195$^{\ddagger}$ & 654866823401111168& 29030$\pm$240& 5.47$\pm$0.04 & -2.96$\pm$0.19 & 31.7 & 25.42 & 14.77 & sdB & \\
123.14029$^{*}$ & 16.023232$^{\dagger}$ & 227309093 & 655674865665718656& 31350$\pm$130& 5.52$\pm$0.02 & -3.00> & -94.2 & 52.05 & 13.69 & sdB & \\
126.89565$^{*}$ & 17.898859$^{\dagger}$ & 268107224 & 662146419308286976& 29460$\pm$200& 5.64$\pm$0.03 & -1.26$\pm$0.03 & -110.8 & 47.66 & 14.58 & sdB & \\
127.136949$^{*}$ & 14.867359$^{\dagger}$ & 380214234 & 651745279826458112& 38230$\pm$150& 5.81$\pm$0.01 & -0.71$\pm$0.02 & -19.3 & 167.39 & 11.64 & He-sdOB & \\
128.551254$^{*}$ & 7.202881$^{\dagger}$ & 340409136 & 595824702553000448& 29550$\pm$130& 5.63$\pm$0.01 & -2.94$\pm$0.11 & -32.1 & 47.46 & 14.68 & sdB & \\
128.849472$^{*}$ & -1.931296 & 183405148 & 3073231760953563264& 46270$\pm$330& 5.88$\pm$0.04 & 0.29$\pm$0.10 & 4.5 & 127.86 & 11.38 & He-sdOB & \\
128.898243 & 19.736805 & 399207128 & 662848247026938240& 27170$\pm$440& 5.41$\pm$0.05 & -2.65$\pm$0.11 & -56.0 & 17.61 & 13.10 & sdB & \\
\hline\noalign{\smallskip} 
  \end{tabularx}
\end{table*}

\setcounter{table}{0}
\begin{table*}
\tiny
 \begin{minipage}{180mm}
  \caption{continued. }
  \end{minipage}\\
  \centering
    \begin{tabularx}{18.0cm}{lllcccccccccccccX}
\hline\noalign{\smallskip}
ra\tablenotemark{a}  & dec\tablenotemark{b}   &obsid\tablenotemark{c} & source\_id 
& $T_\mathrm{eff}$ &  $\mathrm{log}\ g$ & $\mathrm{log}(n\mathrm{He}/n\mathrm{H})$\tablenotemark{d}
& RV    &SNRU &$G$ & spclass\\
 LAMOST &  LAMOST& LAMOST&Gaia  &(K)&($\mathrm{cm\ s^{-2}}$)& & $\mathrm{km\ s^{-1}}$& &Gaia(mag) & \\
\hline\noalign{\smallskip}
130.432816$^{*}$ & 13.075019$^{\dagger}$ & 380207118 & 603097422215268736& 46930$\pm$340& 5.53$\pm$0.06 & 0.16$\pm$0.05 & -34.7 & 92.85 & 13.61 & He-sdOB & \\
130.770989$^{*}$ & 33.462441 & 199614008 & 710337051880367616& 27360$\pm$530& 5.47$\pm$0.03 & -2.80$\pm$0.15 & 95.7 & 19.23 & 14.91 & sdB & \\
130.919521 & 2.007572 & 449410221 & 3078888851357039360& 42780$\pm$1070& 5.28$\pm$0.05 & -2.82$\pm$0.16 & 22.5 & 34.53 & 14.03 & sdO & \\
131.39858 & 19.697389$^{\dagger}$ & 535915112 & 661156102927711104& 23510$\pm$90& 5.07$\pm$0.00 & -1.86$\pm$0.17 & -41.1 & 66.89 & 13.26 & sdB & \\
133.348574$^{*}$ & 16.826456 & 116809080$^{\ddagger}$ & 611684298790301568& 27820$\pm$210& 5.41$\pm$0.03 & -2.77$\pm$0.08 & -38.1 & 18.30 & 13.93 & sdB & \\
134.205286$^{*}$ & 17.020746$^{\dagger}$ & 116812014$^{\ddagger}$ & 611587919724027392& 29080$\pm$420& 5.49$\pm$0.02 & -3.07$\pm$0.18 & 32.8 & 31.59 & 12.81 & sdB & \\
134.71299$^{*}$ & 2.170267 & 109804103 & 577763368640517760& 48090$\pm$1480& 5.62$\pm$0.06 & -1.93$\pm$0.14 & -46.5 & 20.17 & 13.63 & sdO & \\
134.761016 & 11.94105 & 93916015$^{\ddagger}$ & 604794342318395520& 25770$\pm$360& 5.61$\pm$0.03 & -3.00> & -100.7 & 12.94 & 13.47 & sdB & \\
136.4204$^{*}$ & 12.207931$^{\dagger}$ & 531216109 & 604474251290699776& 28570$\pm$110& 5.47$\pm$0.03 & -2.62$\pm$0.07 & 12.4 & 46.95 & 14.62 & sdB & \\
137.605977$^{*}$ & 12.140855 & 94009080$^{\ddagger}$ & 604238612204857088& 27430$\pm$110& 5.36$\pm$0.01 & -2.82$\pm$0.05 & 22.2 & 30.75 & 13.93 & sdB & \\
138.03039$^{*}$ & 16.222359 & 200801161$^{\ddagger}$ & 607898538520616704& 39490$\pm$990& 5.37$\pm$0.04 & -3.02$\pm$0.22 & -49.2 & 89.45 & 13.81 & sdO & \\
138.215253$^{*}$ & 27.342056 & 44602153$^{\ddagger}$ & 695059062934110592& 36700$\pm$120& 5.96$\pm$0.04 & -0.83$\pm$0.04 & -14.5 & 48.12 & 12.21 & He-sdOB & \\
140.36756$^{*}$ & 2.767307 & 101704108$^{\ddagger}$ & 3845583703883921920& 31610$\pm$240& 5.84$\pm$0.04 & -2.46$\pm$0.10 & -50.4 & 26.27 & 13.30 & sdB & \\
140.665973$^{*}$ & 27.040402 & 100508062$^{\ddagger}$ & 694109462844643072& 31770$\pm$70& 5.89$\pm$0.01 & -2.51$\pm$0.08 & 32.4 & 29.49 & 12.62 & sdB & \\
140.805888$^{*}$ & 29.449328 & 11307210$^{\ddagger}$ & 695726844449223040& 30870$\pm$300& 6.24$\pm$0.07 & -1.34$\pm$0.06 & -56.4 & 16.65 & 14.71 & sdOB & \\
142.233763$^{*}$ & 6.276275$^{\dagger}$ & 223707059 & 586043309672965888& 27240$\pm$150& 5.47$\pm$0.01 & -2.80$\pm$0.07 & 30.4 & 36.65 & 14.10 & sdB & \\
142.564419$^{*}$ & 30.842911 & 556014133$^{\ddagger}$ & 697406932576439680& 29100$\pm$160& 5.67$\pm$0.02 & -2.34$\pm$0.05 & -113.8 & 48.05 & 14.88 & sdB & \\
142.74847$^{*}$ & 2.842343 & 206408094 & 3844661557225496064& 31340$\pm$210& 5.58$\pm$0.02 & -2.56$\pm$0.07 & -55.4 & 29.48 & 14.95 & sdB & \\
144.543089$^{*}$ & -3.999546 & 408905050 & 3824642199263291136& 34410$\pm$290& 5.87$\pm$0.03 & -1.82$\pm$0.06 & -55.3 & 27.95 & 14.68 & sdOB & \\
146.299391$^{*}$ & -3.155844 & 408906081 & 3826259821385575296& 40330$\pm$510& 6.24$\pm$0.03 & -3.00> & 31.3 & 66.40 & 13.97 & sdO & \\
149.884416$^{*}$ & 36.30718 & 90914031$^{\ddagger}$ & 795959630108543232& 27620$\pm$180& 5.22$\pm$0.01 & -2.70$\pm$0.30 & 32.4 & 65.43 & 12.82 & sdB & \\
150.976151$^{*}$ & 40.571695 & 99507015$^{\ddagger}$ & 803736166614876544& 41260$\pm$850& 5.33$\pm$0.04 & -2.88$\pm$0.18 & 22.4 & 52.52 & 13.27 & sdO & \\
159.90309$^{*}$ & 43.102566 & 228703057 & 781164326766404736& 71950$\pm$5110& 5.67$\pm$0.03 & -1.74$\pm$0.04 & 30.3 & 98.07 & 11.11 & sdO & \\
162.511777$^{*}$ & -0.010237 & 205211114 & 3806303066866089216& 34850$\pm$90& 5.83$\pm$0.01 & -1.89$\pm$0.06 & 13.6 & 49.68 & 13.43 & sdOB & \\
165.973167$^{*}$ & 11.001214 & 290015051 & 3868458111291375616& 35750$\pm$1050& 5.87$\pm$0.34 & -1.55$\pm$0.30 & 5.4 & 27.11 & 15.12 & sdOB & \\
169.205762 & 6.992267 & 401216228 & 3817717887347994112& 30570$\pm$30& 5.60$\pm$0.01 & -2.52$\pm$0.05 & 16.3 & 45.34 & 13.00 & sdB & \\
170.001536$^{*}$ & 30.950015 & 300414223 & 4023386553845346688& 33620$\pm$350& 5.77$\pm$0.04 & -2.33$\pm$0.08 & 10.8 & 34.28 & 14.51 & sdOB & \\
170.961175$^{*}$ & 23.612733 & 309515234 & 3992841536709888640& 28080$\pm$60& 5.38$\pm$0.01 & -2.36$\pm$0.03 & -37.6 & 67.31 & 14.15 & sdB & \\
172.51545$^{*}$ & 1.627016 & 237307119$^{\ddagger}$ & 3799987433421443712& 42390$\pm$300& 5.68$\pm$0.07 & 0.78$\pm$0.07 & -23.0 & 36.42 & 13.79 & He-sdOB & \\
181.601508$^{*}$ & 57.159922$^{\dagger}$ & 25714089$^{\ddagger}$ & 1575432697534110336& 35220$\pm$350& 5.75$\pm$0.04 & -1.75$\pm$0.06 & 80.6 & 22.21 & 14.85 & sdOB & \\
188.963098$^{*}$ & 42.377703 & 133410210$^{\ddagger}$ & 1534536701840624768& 26750$\pm$220& 5.51$\pm$0.06 & -2.56$\pm$0.07 & -38.5 & 30.67 & 11.99 & sdB & \\
189.219429$^{*}$ & 50.253856 & 22713239$^{\ddagger}$ & 1568451554613413888& 48380$\pm$1350& 5.23$\pm$0.06 & -2.32$\pm$0.12 & 92.6 & 21.98 & 14.65 & sdO & \\
189.347983$^{*}$ & 25.06663 & 555614008 & 3959631234670040704& 33220$\pm$370& 6.14$\pm$0.08 & -1.61$\pm$0.11 & 6.5 & 12.28 & 10.47 & sdOB & \\
190.159493$^{*}$ & 51.266733 & 455403016 & 1568569889552108800& 42190$\pm$180& 5.72$\pm$0.05 & 0.69$\pm$0.01 & 64.9 & 77.38 & 13.58 & He-sdOB & \\
190.629575$^{*}$ & 4.112153 & 206508081 & 3704162319296853120& 34180$\pm$60& 5.87$\pm$0.02 & -2.04$\pm$0.03 & -49.2 & 28.42 & 15.48 & sdOB & \\
192.709449$^{*}$ & 16.167535 & 100802163$^{\ddagger}$ & 3934486949931442176& 26540$\pm$690& 5.53$\pm$0.05 & -2.47$\pm$0.10 & -37.1 & 18.93 & 14.41 & sdB & \\
194.455489$^{*}$ & 54.42649 & 415303223 & 1570270249924113408& 33130$\pm$150& 5.89$\pm$0.02 & -1.55$\pm$0.04 & 4.3 & 49.83 & 13.50 & sdOB & \\
194.838437$^{*}$ & 27.568081 & 301804130 & 1460692714941449728& 35550$\pm$590& 5.73$\pm$0.08 & -0.79$\pm$0.07 & -65.6 & 12.95 & 14.19 & He-sdOB & \\
195.67403$^{*}$ & 27.678316 & 235908240 & 1460730407575030656& 38220$\pm$8090& 6.74$\pm$0.82 & -2.07 & 407.3 & 10.97 & 14.21 & sdB & \\
196.516153$^{*}$ & 9.408616 & 342107122 & 3734224925705249152& 29870$\pm$250& 5.76$\pm$0.02 & -3.35$\pm$0.23 & -33.5 & 35.04 & 14.45 & sdB & \\
198.942$^{*}$ & 16.224036 & 424903052 & 3744601257454417152& 32070$\pm$150& 5.85$\pm$0.05 & -2.04$\pm$0.05 & 16.3 & 45.27 & 15.79 & sdOB & \\
202.973147$^{*}$ & 15.688212 & 142716122$^{\ddagger}$ & 3744968013301724544& 29370$\pm$250& 5.53$\pm$0.04 & -2.87$\pm$0.28 & 33.1 & 22.30 & 13.50 & sdB & \\
204.011629 & 36.091513 & 338308237 & 1471988375850922368& 34250$\pm$300& 5.91$\pm$0.05 & -1.68$\pm$0.06 & 18.5 & 37.81 & 14.65 & sdOB & \\
204.70065$^{*}$ & -2.030358 & 212806164 & 3637481302758519040& 32500$\pm$50& 5.79$\pm$0.01 & -2.91$\pm$0.08 & 84.9 & 131.58 & 13.39 & sdB & \\
205.036798$^{*}$ & 47.864419 & 143413093$^{\ddagger}$ & 1552221800914196480& 28480$\pm$230& 5.40$\pm$0.03 & -2.70$\pm$0.10 & 27.5 & 28.45 & 13.58 & sdB & \\
212.567679$^{*}$ & 32.412925 & 230113142 & 1478424229724415104& 31620$\pm$340& 6.03$\pm$0.06 & -2.17$\pm$0.16 & -36.7 & 12.06 & 14.90 & sdB & \\
212.607982$^{*}$ & -1.504604 & 449815203 & 3647329074356945152& 36690$\pm$500& 5.58$\pm$0.01 & -2.85$\pm$0.12 & 99.3 & 46.77 & 13.72 & sdOB & \\
213.764081$^{*}$ & 41.8243 & 575416138 & 1492192692285509376& 30910$\pm$230& 5.64$\pm$0.07 & -2.93$\pm$0.23 & 35.3 & 27.02 & 14.79 & sdB & \\
216.283123$^{*}$ & 20.378427 & 316812154 & 1240306604269576704& 35660$\pm$450& 5.72$\pm$0.08 & -1.56$\pm$0.09 & 8.8 & 34.12 & 14.98 & sdOB & \\
217.123294$^{*}$ & 21.10059 & 307805237 & 1240449403342559360& 32550$\pm$80& 5.86$\pm$0.03 & -1.62$\pm$0.02 & 31.5 & 50.50 & 13.22 & sdB & \\
218.828893$^{*}$ & 15.670621 & 418712144 & 1234322477875249664& 26210$\pm$210& 5.47$\pm$0.05 & -2.52$\pm$0.10 & 95.1 & 22.46 & 13.86 & sdB & \\
220.220106$^{*}$ & -3.147965 & 132203002$^{\ddagger}$ & 3648191164488514688& 29490$\pm$260& 5.45$\pm$0.03 & -2.90$\pm$0.10 & 33.8 & 43.93 & 13.82 & sdB & \\
220.322231$^{*}$ & 1.624409 & 527203236 & 3655309299687364992& 29110$\pm$300& 5.75$\pm$0.02 & -2.38$\pm$0.11 & 30.5 & 53.88 & 14.26 & sdB & \\
223.665896$^{*}$ & 19.616899 & 234115038 & 1237226700401153792& 31000$\pm$140& 5.88$\pm$0.02 & -2.37$\pm$0.03 & -55.7 & 64.13 & 12.45 & sdB & \\
225.102211$^{*}$ & 42.095753 & 580111192 & 1585277243613047936& 31000$\pm$130& 5.59$\pm$0.03 & -3.00> & 99.7 & 105.90 & 13.88 & sdB & \\
230.1914$^{*}$ & 29.80715 & 219416182 & 1275262415380252544& 37890$\pm$480& 6.03$\pm$0.07 & -1.21$\pm$0.08 & 160.9 & 13.06 & 15.68 & sdOB & \\
232.708055$^{*}$ & 6.015569 & 566509196 & 4429114659552190080& 33840$\pm$80& 6.00$\pm$0.02 & -1.45$\pm$0.03 & 15.2 & 67.63 & 14.71 & sdOB & \\
232.762545 & 33.638114 & 333013143 & 1373881186687373696& 31680$\pm$190& 5.92$\pm$0.04 & -1.63$\pm$0.04 & 13.5 & 34.76 & 14.78 & sdB & \\
232.79329$^{*}$ & 10.250305 & 243312195 & 1165963744033018496& 26800$\pm$870& 5.51$\pm$0.08 & -2.47$\pm$0.11 & 24.2 & 19.02 & 13.57 & sdB & \\
233.317108$^{*}$ & 44.587383 & 582404042 & 1394581623383963264& 30400$\pm$150& 5.80$\pm$0.03 & -1.91$\pm$0.03 & -107.0 & 56.07 & 15.54 & sdB & \\
233.456025$^{*}$ & 37.991129 & 331509102 & 1375814952762454272& 29300$\pm$120& 5.55$\pm$0.01 & -2.34$\pm$0.03 & 79.0 & 78.16 & 12.98 & sdB & \\
233.815317$^{*}$ & 34.790072 & 322207235 & 1374058963973357696& 38230$\pm$420& 5.57$\pm$0.05 & -3.44$\pm$0.34 & 17.3 & 26.53 & 15.05 & sdB & \\
235.097774$^{*}$ & 26.808252 & 328611040 & 1223920926079402112& 26220$\pm$120& 5.41$\pm$0.04 & -3.23$\pm$0.10 & 29.0 & 38.01 & 13.82 & sdB & \\
235.162647$^{*}$ & 39.930281 & 152406202 & 1377891689708857856& 33520$\pm$650& 6.25$\pm$0.13 & -3.00> & 8.5 & 11.94 & 13.21 & sdB & \\
236.548706$^{*}$ & 48.643679 & 155111084$^{\ddagger}$ & 1401621384019650560& 40900$\pm$320& 6.27$\pm$0.07 & 4.01$\pm$0.06 & 70.1 & 19.56 & 12.79 & He-sdB & \\
237.701252$^{*}$ & 16.440384 & 571812247 & 1196375273586921472& 32050$\pm$60& 5.68$\pm$0.04 & -2.06$\pm$0.04 & 10.0 & 44.48 & 14.85 & sdOB & \\
240.300504$^{*}$ & 53.197761 & 149514028$^{\ddagger}$ & 1404827731724827776& 30940$\pm$170& 5.79$\pm$0.03 & -2.44$\pm$0.06 & 86.4 & 31.64 & 14.25 & sdB & \\
240.380311$^{*}$ & 4.674167$^{\dagger}$ & 133105185$^{\ddagger}$ & 4425487989169416320& 38100$\pm$360& 5.74$\pm$0.05 & -0.38$\pm$0.02 & -0.2 & 11.42 & 14.48 & He-sdOB & \\
242.015365$^{*}$ & 7.074661 & 133011059$^{\ddagger}$ & 4450489165598124800& 32480$\pm$270& 5.34$\pm$0.03 & -2.54$\pm$0.09 & 32.7 & 26.69 & 12.83 & sdB & \\
242.478874 & 17.249217 & 447906246 & 1198927755470384128& 29170$\pm$70& 5.12$\pm$0.02 & -0.47$\pm$0.03 & -1.5 & 159.59 & 12.22 & He-sdOB & \\
242.838926$^{*}$ & 52.768252 & 332203009 & 1427969118595251072& 30870$\pm$190& 5.66$\pm$0.04 & -2.23$\pm$0.14 & -35.8 & 78.27 & 12.77 & sdB & \\
243.002746$^{*}$ & 51.82875 & 149505207$^{\ddagger}$ & 1427678435208512384& 34620$\pm$380& 5.08$\pm$0.05 & -3.00> & -37.4 & 25.11 & 13.67 & sdOB & \\
244.10151$^{*}$ & 26.88743 & 574101201 & 1315765159572132224& 44040$\pm$20& 6.66$\pm$0.01 & 3.61$\pm$0.04 & 60.4 & 37.24 & 15.90 & He-sdB & \\
244.596204$^{*}$ & 14.270251 & 337915062 & 4463625890008114944& 36830$\pm$60& 6.07$\pm$0.01 & -1.45$\pm$0.02 & 17.0 & 58.18 & 13.44 & sdOB & \\
244.84679$^{*}$ & 30.833913 & 220412139 & 1319053146015474688& 36610$\pm$790& 5.98$\pm$0.12 & -1.50$\pm$0.12 & 11.6 & 16.18 & 15.63 & sdOB & \\
246.132846$^{*}$ & 40.986953$^{\dagger}$ & 241901157 & 1380974582874497280& 28950$\pm$120& 5.42$\pm$0.01 & -2.48$\pm$0.04 & 28.4 & 87.55 & 14.46 & sdB & \\
246.818754$^{*}$ & 40.457908$^{\dagger}$ & 576301124 & 1332896306646572160& 23710$\pm$60& 5.49$\pm$0.01 & -2.93$\pm$0.05 & 36.0 & 228.91 & 12.56 & sdOB & \\
247.093844$^{*}$ & 27.488591$^{\dagger}$ & 225003186 & 1305126525580012672& 27750$\pm$180& 5.42$\pm$0.03 & -2.52$\pm$0.03 & -40.3 & 107.87 & 13.64 & sdB & \\
248.005651$^{*}$ & 7.994445$^{\dagger}$ & 140315205 & 4440145853560129408& 38450$\pm$840& 5.42$\pm$0.01 & -2.71$\pm$0.04 & -47.4 & 63.05 & 12.87 & sdO & \\
249.26495$^{*}$ & 5.984855 & 571908139 & 4435849168273656704& 30750$\pm$130& 5.50$\pm$0.04 & -2.42$\pm$0.15 & 32.3 & 66.60 & 13.84 & sdB & \\
\hline\noalign{\smallskip} 
  \end{tabularx}
\end{table*}

\setcounter{table}{0}
\begin{table*}
\tiny

 \begin{minipage}{180mm}
  \caption{continued. }
  \end{minipage}\\
  \centering
    \begin{tabularx}{18.0cm}{lllcccccccccccccX}
\hline\noalign{\smallskip}
ra\tablenotemark{a}  & dec\tablenotemark{b}   &obsid\tablenotemark{c} & source\_id 
& $T_\mathrm{eff}$ &  $\mathrm{log}\ g$ & $\mathrm{log}(n\mathrm{He}/n\mathrm{H})$\tablenotemark{d}
& RV   &SNRU &$G$ & spclass\\
 LAMOST &  LAMOST& LAMOST&Gaia  &(K)&($\mathrm{cm\ s^{-2}}$)& & $\mathrm{km\ s^{-1}}$& &Gaia(mag) & \\
\hline\noalign{\smallskip}
251.154843$^{*}$ & -0.001554 & 340308108 & 4382904797178453504& 26430$\pm$290& 5.51$\pm$0.04 & -2.59$\pm$0.06 & 88.3 & 25.03 & 13.66 & sdB & \\
252.2449$^{*}$ & 24.967189 & 567901136 & 1300121926607765760& 23390$\pm$370& 5.72$\pm$0.04 & -3.00> & 27.0 & 17.52 & 16.07 & sdB & \\
252.499386$^{*}$ & 53.525487 & 148415067$^{\ddagger}$ & 1425987175870116864& 31000$\pm$740& 5.60$\pm$0.12 & -3.00> & 35.5 & 15.11 & 14.02 & sdB & \\
253.928415$^{*}$ & 13.030146 & 144307163 & 4449162845336419840& 26230$\pm$550& 5.34$\pm$0.06 & -2.50$\pm$0.10 & 17.3 & 25.82 & 14.38 & sdB & \\
257.3859$^{*}$ & 21.370697 & 575513248 & 4567834303554972288& 35690$\pm$280& 5.80$\pm$0.05 & -1.40$\pm$0.05 & 14.6 & 23.45 & 15.61 & sdOB & \\
257.555047 & 53.446121 & 148413001 & 1419133610657016064& 24010$\pm$440& 5.54$\pm$0.11 & -2.13$\pm$0.11 & -55.5 & 14.11 & 12.60 & sdB & \\
258.078166$^{*}$ & 48.976605 & 148506234$^{\ddagger}$ & 1414187320160273024& 29870$\pm$170& 5.85$\pm$0.05 & -2.45$\pm$0.12 & 31.3 & 37.53 & 12.85 & sdB & \\
259.32832 & 42.435913 & 141008222 & 1360139937040515072& 66410$\pm$10150& 5.67$\pm$0.07 & -2.88$\pm$0.43 & 3.7 & 31.96 & 12.48 & sdO & \\
260.35745$^{*}$ & 9.053239 & 241603163 & 4491274930955326080& 26640$\pm$490& 5.55$\pm$0.04 & -2.95$\pm$0.17 & 32.6 & 19.63 & 15.62 & sdB & \\
261.688678$^{*}$ & 37.225178 & 462604059 & 1337242538672314112& 36250$\pm$3050& 5.86$\pm$0.47 & -1.28$\pm$0.28 & 7.1 & 23.01 & 15.51 & sdOB & \\
261.989091$^{*}$ & 25.143254 & 584512022 & 4570629777508086400& 27540$\pm$390& 5.47$\pm$0.07 & -3.00> & 106.3 & 11.43 & 12.99 & sdB & \\
264.126193$^{*}$ & 7.600641$^{\dagger}$ & 341003194 & 4487346891305390592& 27890$\pm$200& 5.37$\pm$0.02 & -3.14$\pm$0.12 & 56.5 & 26.28 & 13.70 & sdB & \\
264.360098$^{*}$ & 22.149371 & 446111200 & 4557081835745623808& 41170$\pm$590& 5.28$\pm$0.02 & -2.97$\pm$0.07 & 84.8 & 154.48 & 11.82 & sdO & \\
264.776732 & 11.296711 & 242308023 & 4492802221326659456& 28430$\pm$160& 5.52$\pm$0.02 & -2.80$\pm$0.06 & -107.9 & 67.02 & 13.30 & sdB & \\
265.091562$^{*}$ & 48.899451 & 580907209 & 1363886900869046656& 27550$\pm$290& 5.41$\pm$0.03 & -2.70$\pm$0.07 & 63.6 & 44.29 & 13.19 & sdB & \\
265.82931$^{*}$ & 21.543762 & 457610044 & 4556063035156496768& 37590$\pm$630& 5.56$\pm$0.08 & -2.42$\pm$0.13 & 153.8 & 16.01 & 14.01 & sdOB & \\
266.726507$^{*}$ & 43.76783 & 568415057 & 1349088676950060672& 33750$\pm$250& 5.67$\pm$0.01 & -1.66$\pm$0.04 & 13.7 & 67.95 & 14.38 & sdOB & \\
267.483551 & 0.110035 & 459005109 & 4371689194462064768& 32600$\pm$100& 5.89$\pm$0.01 & -1.81$\pm$0.06 & 32.8 & 23.75 & 14.36 & sdOB & \\
268.270067 & 24.596209 & 457611042 & 4581048921494269056& 37320$\pm$540& 5.40$\pm$0.05 & -3.07$\pm$0.26 & 10.2 & 18.09 & 14.61 & sdB & \\
272.388874 & 22.516659 & 234607117 & 4577984720030433536& 27480$\pm$50& 5.53$\pm$0.01 & -2.62$\pm$0.03 & 29.2 & 43.85 & 12.94 & sdB & \\
279.443598$^{*}$ & 59.26635 & 346815163 & 2154810316748450432& 39210$\pm$70& 5.72$\pm$0.05 & -0.17$\pm$0.03 & 11.9 & 15.24 & 14.59 & He-sdOB & \\
285.754739 & 34.758065 & 574208088 & 2092525388221505792& 28440$\pm$190& 5.45$\pm$0.02 & -2.97$\pm$0.05 & 33.3 & 47.26 & 15.69 & sdB & \\
285.90424$^{*}$ & 38.603512 & 576405016 & 2100027626635533056& 29050$\pm$120& 5.39$\pm$0.03 & -2.88$\pm$0.07 & 14.2 & 48.23 & 14.66 & sdB & \\
286.918867$^{*}$ & 42.306115 & 161401152 & 2105469320138052864& 30560$\pm$260& 5.10$\pm$0.04 & -1.59$\pm$0.04 & 9.2 & 136.24 & 10.59 & sdB & \\
287.27975$^{*}$ & 37.937321 & 576407250 & 2099149567821051904& 24320$\pm$90& 5.42$\pm$0.02 & -2.78$\pm$0.06 & 29.2 & 36.17 & 15.26 & sdB & \\
288.815187 & 43.674379 & 362008075 & 2103044549103238912& 27850$\pm$240& 5.57$\pm$0.04 & -2.55$\pm$0.04 & 19.7 & 79.13 & 12.46 & sdB & \\
291.52478 & 47.275349 & 354204250 & 2129130844728397568& 62190$\pm$1470& 5.65$\pm$0.07 & 0.00$\pm$0.04 & 26.2 & 62.45 & 14.86 & He-sdO & \\
291.539417$^{*}$ & 37.335611 & 52401101 & 2051770507972278016& 30670$\pm$160& 5.82$\pm$0.05 & -1.61$\pm$0.04 & 2.6 & 20.45 & 13.61 & sdB & \\
291.64212$^{*}$ & 49.50824 & 353808152 & 2129777323205768448& 27340$\pm$1140& 5.56$\pm$0.13 & -3.00> & 94.2 & 13.85 & 14.80 & sdB & \\
294.635917$^{*}$ & 46.066417 & 247615029 & 2080063931448749824& 30030$\pm$80& 5.43$\pm$0.02 & -2.18$\pm$0.06 & -37.3 & 109.98 & 12.14 & sdB & \\
294.82629 & 47.14856 & 354605098 & 2128554563197280512& 29820$\pm$170& 5.42$\pm$0.04 & -2.32$\pm$0.10 & 30.9 & 19.83 & 15.40 & sdB & \\
296.05313 & 50.494251 & 354611122 & 2135353702585217920& 27770$\pm$280& 5.50$\pm$0.05 & -2.31$\pm$0.08 & 149.9 & 21.90 & 15.48 & sdB & \\
297.984417$^{*}$ & 46.130778 & 165903109 & 2079522048319286528& 39380$\pm$420& 5.51$\pm$0.03 & -3.27$\pm$0.51 & 78.2 & 50.64 & 13.63 & sdO & \\
298.28497 & 47.716728 & 354606155 & 2086441927825846656& 28810$\pm$300& 5.42$\pm$0.05 & -2.71$\pm$0.11 & 30.9 & 34.92 & 15.00 & sdB & \\
299.231536 & 43.837941 & 355101074 & 2079009641541181312& 39060$\pm$180& 5.72$\pm$0.04 & -0.83$\pm$0.05 & 21.1 & 106.81 & 14.34 & He-sdOB & \\
300.179568 & 23.267099 & 577014004 & 1833281127895263872& 35410$\pm$310& 5.93$\pm$0.03 & -1.45$\pm$0.03 & 77.0 & 74.30 & 11.97 & sdOB & \\
300.39346 & 44.574322 & 355208177 & 2076135724307679872& 34260$\pm$1220& 5.29$\pm$0.16 & -2.75 & -4.4 & 38.30 & 15.45 & sdB & \\
301.72862$^{*}$ & 14.715003 & 462109242 & 1806906627074342016& 34620$\pm$480& 6.08$\pm$0.10 & -1.60$\pm$0.09 & 14.7 & 10.48 & 14.34 & sdOB & \\
311.28078 & 36.636963 & 470909040 & 1870606730148932224& 25040$\pm$180& 5.79$\pm$0.04 & -3.00> & 37.1 & 15.36 & 16.48 & sdB & \\
312.413176 & 30.081818 & 475203139 & 1858674589447492608& 37330$\pm$170& 5.54$\pm$0.06 & -2.58$\pm$0.12 & 20.2 & 98.43 & 13.50 & sdOB & \\
315.580451$^{*}$ & 1.537663 & 370205063 & 1729653290822954880& 34480$\pm$300& 5.73$\pm$0.07 & -1.43$\pm$0.05 & 8.3 & 47.99 & 15.00 & sdOB & \\
315.885216$^{*}$ & 30.593842 & 475213199 & 1852616039853485824& 34280$\pm$160& 5.90$\pm$0.03 & -1.77$\pm$0.03 & 10.6 & 63.06 & 12.95 & sdOB & \\
317.372429 & 13.279525 & 371607236 & 1758993857393963776& 26520$\pm$450& 5.52$\pm$0.06 & -2.62$\pm$0.11 & -38.0 & 25.36 & 15.42 & sdB & \\
317.672454$^{*}$ & 14.433031$^{\dagger}$ & 371606237 & 1759968097711823232& 43210$\pm$270& 5.83$\pm$0.04 & 0.56$\pm$0.04 & 45.0 & 106.13 & 14.19 & He-sdOB & \\
318.417925$^{*}$ & 20.715492 & 372014020 & 1789840488608448640& 26520$\pm$310& 5.29$\pm$0.03 & -2.58$\pm$0.08 & 100.5 & 15.55 & 15.08 & sdB & \\
321.435002$^{*}$ & 20.469909$^{\dagger}$ & 503309030 & 1790336673293954176& 50130$\pm$380& 5.57$\pm$0.04 & 0.20$\pm$0.04 & 5.2 & 64.56 & 13.37 & He-sdO & \\
322.335687 & 29.968578 & 368208087 & 1850084998438797312& 34710$\pm$180& 6.17$\pm$0.03 & -1.76$\pm$0.04 & 79.5 & 70.36 & 14.10 & sdOB & \\
322.871071 & 19.865855 & 503313058 & 1787066145893240576& 39960$\pm$710& 5.94$\pm$0.02 & -2.80$\pm$0.17 & 11.4 & 24.68 & 13.16 & sdO & \\
324.5772 & 34.351566 & 472809107 & 1950263255195349888& 30340$\pm$180& 5.41$\pm$0.03 & -1.82$\pm$0.05 & 11.8 & 34.39 & 14.74 & sdB & \\
324.613116 & 46.192433 & 476303135 & 1977291346957976320& 29330$\pm$300& 5.80$\pm$0.09 & -1.86$\pm$0.08 & 10.1 & 23.71 & 16.16 & sdOB & \\
324.833471 & 39.546916 & 172101139 & 1953965379566626816& 46900$\pm$320& 5.61$\pm$0.08 & 0.10$\pm$0.04 & 3.7 & 60.81 & 14.73 & He-sdOB & \\
325.278181 & 25.474982 & 471009116 & 1797986937760585600& 43540$\pm$240& 5.91$\pm$0.06 & 0.52$\pm$0.04 & 0.2 & 31.87 & 14.11 & He-sdOB & \\
325.676688 & 33.081334 & 472808084 & 1946937541760735872& 35950$\pm$550& 5.69$\pm$0.11 & -1.36$\pm$0.09 & 14.0 & 39.62 & 14.79 & sdOB & \\
326.501298 & 37.35549 & 473209201 & 1952553606634620928& 51430$\pm$360& 5.74$\pm$0.08 & 0.16$\pm$0.05 & 74.4 & 31.05 & 14.66 & He-sdO & \\
330.2594$^{*}$ & 8.513191 & 256811238 & 2724882872133120512& 65560$\pm$1450& 5.83$\pm$0.04 & 0.06$\pm$0.05 & 99.5 & 70.70 & 13.13 & He-sdO & \\
332.511298 & 25.066193 & 164209157 & 1879544900675463552& 21290$\pm$80& 5.32$\pm$0.02 & -2.39$\pm$0.04 & 32.9 & 71.00 & 12.83 & sdB & \\
339.4026$^{*}$ & 22.73709 & 490614158 & 1874534323107953664& 42390$\pm$890& 5.40$\pm$0.03 & -3.09$\pm$0.10 & 31.3 & 83.68 & 12.76 & sdO & \\
340.718491$^{*}$ & 1.872661 & 362211238 & 2654793575707334784& 31140$\pm$90& 5.93$\pm$0.02 & -2.36$\pm$0.04 & -108.9 & 58.30 & 14.44 & sdB & \\
340.871872$^{*}$ & 10.781563 & 157006179 & 2717582699040219392& 34270$\pm$180& 5.77$\pm$0.04 & -1.57$\pm$0.04 & 16.2 & 87.75 & 15.04 & sdOB & \\
341.890375$^{*}$ & 31.325914 & 264909113 & 1889631201873127424& 28940$\pm$240& 5.55$\pm$0.04 & -2.63$\pm$0.12 & 29.4 & 43.26 & 15.45 & sdB & \\
343.992447$^{*}$ & 33.719939 & 370916040 & 1891098500140100352& 28650$\pm$270& 5.64$\pm$0.07 & -2.86$\pm$0.17 & 35.1 & 12.05 & 12.77 & sdB & \\
344.15977$^{*}$ & 6.947541 & 256106208 & 2712047826225618048& 29920$\pm$270& 5.65$\pm$0.05 & -2.79$\pm$0.13 & -40.4 & 42.85 & 15.26 & sdB & \\
345.4409$^{*}$ & 13.64374 & 363809200 & 2815153223450428032& 31190$\pm$220& 5.73$\pm$0.04 & -1.74$\pm$0.04 & 26.5 & 79.42 & 14.52 & sdB & \\
346.401764$^{*}$ & 34.698363 & 164306246 & 1914301803258669440& 28260$\pm$160& 5.50$\pm$0.01 & -2.34$\pm$0.03 & 14.2 & 72.89 & 13.29 & sdB & \\
347.375115$^{*}$ & 28.568276 & 470009143 & 2845710850610464640& 31890$\pm$160& 5.73$\pm$0.03 & -1.72$\pm$0.03 & 12.9 & 76.25 & 14.48 & sdOB & \\
349.980633$^{*}$ & 4.876289 & 56601022 & 2660835289023309696& 41390$\pm$2960& 6.23$\pm$0.23 & -3.24$\pm$1.06 & 143.2 & 11.20 & 12.81 & sdB & \\
351.114427$^{*}$ & 21.647578 & 260216070 & 2837982520817733376& 41900$\pm$40& 6.05$\pm$0.04 & 0.98$\pm$0.06 & 42.0 & 48.53 & 13.74 & He-sdOB & \\
355.0197$^{*}$ & 7.285939 & 361504161 & 2757157489778320000& 30720$\pm$190& 5.93$\pm$0.07 & -1.88$\pm$0.04 & 15.9 & 95.88 & 13.47 & sdB & \\
358.822256$^{*}$ & 18.337599 & 54914220$^{\ddagger}$ & 2774038154361038976& 47080$\pm$950& 5.44$\pm$0.13 & 0.53$\pm$0.10 & 2.2 & 21.80 & 13.32 & He-sdOB & \\
\hline\noalign{\smallskip} 
  \end{tabularx}
\end{table*}

\subsection{Comparison with other studies}
\begin{figure}
\centering
\includegraphics [width=120mm]{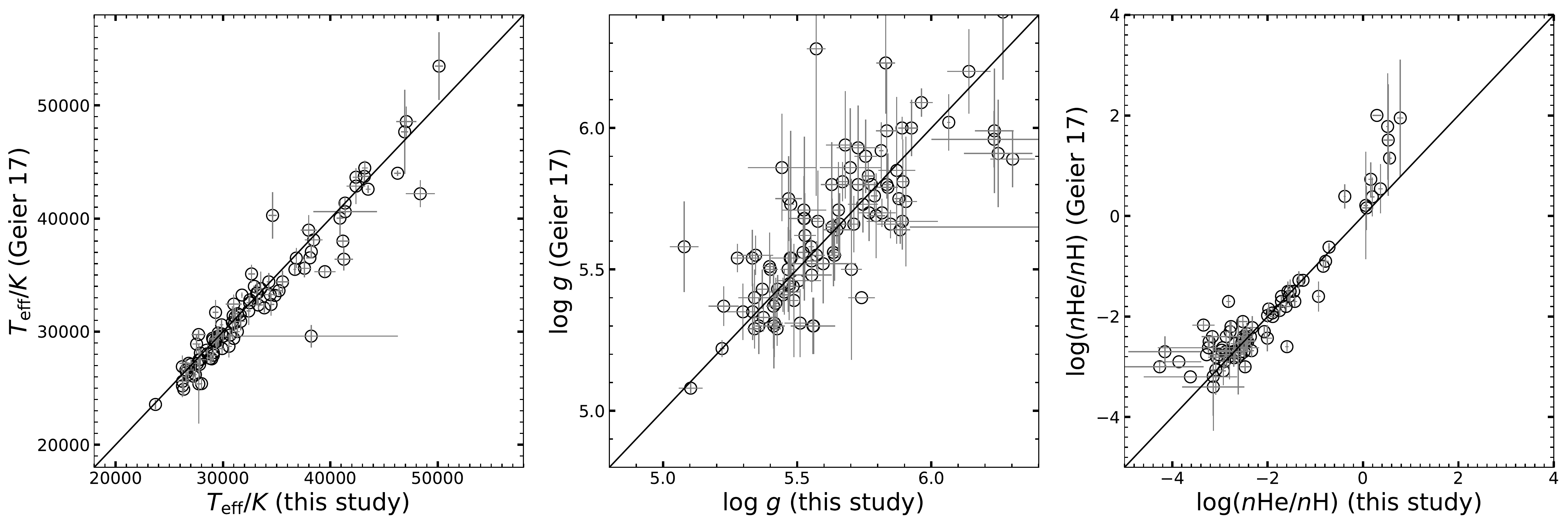}
\caption{Comparisons of atmospheric parameters between this study and 
Geier et al. (2017).   }
\end{figure}

\begin{figure}
\centering
\includegraphics [width=120mm]{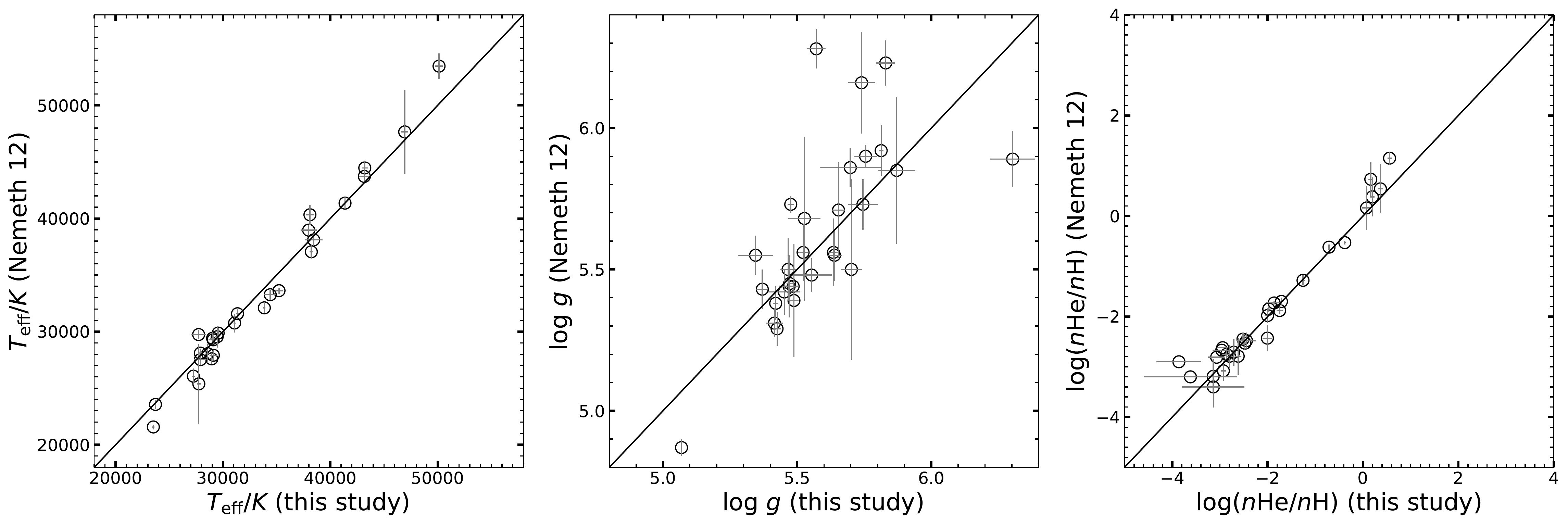}
\caption{Comparisons of atmospheric parameters between this study and 
N\'emeth et al. (2012). }
\end{figure}

\begin{figure}
\centering
\includegraphics [width=120mm]{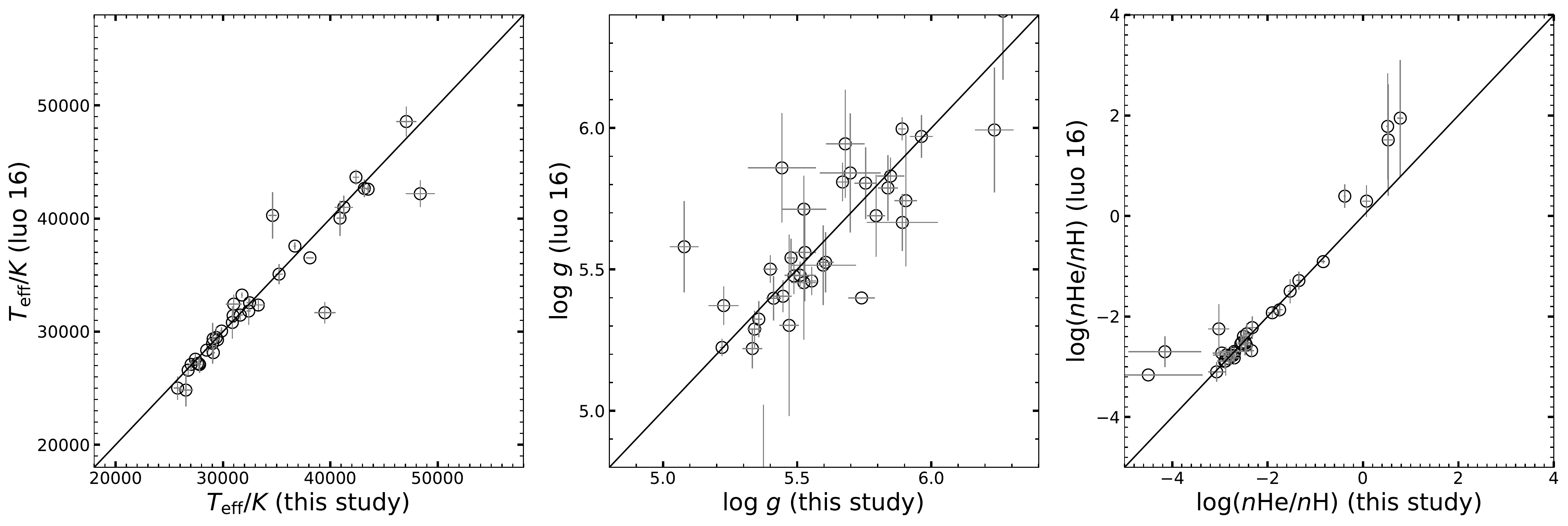}
\caption{Comparisons of atmospheric parameters between this study and 
Luo et al. (2016). }
\end{figure}

Fig 3 shows the comparison of atmospheric parameters for the common stars 
between this study and Geier et al. (2017), for which atmospheric parameters are available. 
One can see that both $T_\mathrm{eff}$ (left panel) and 
$\mathrm{log}(n\mathrm{He}/n\mathrm{H})$ (right panel) 
agree well. Although the comparison of $\mathrm{log}\,g$ (middle panel) 
presents a litter larger dispersion than the other two parameters, 
our results are comparable with the values from Geier et al. (2017).  
Fig 4 and Fig 5 show the comparison between the results from our 
study and the results by N\'emeth et al. (2012) and Luo et al. (2016), respectively. 
Similarly, the values of $T_\mathrm{eff}$ (left panel) and 
$\mathrm{log}(n\mathrm{He}/n\mathrm{H})$ (right panel) obtained in our study 
are very similar to the values  obtained in N\'emeth et al. (2012) and Luo et al. (2016) 
for the same stars. The value of $\mathrm{log}\,g$ (middle panel) are 
also comparable between these studies. Note that the atmospheric parameters 
in N\'emeth et al. (2012)  are obtained by employing {\sc XTgrid} as well. 
The synthetic spectra in N\'emeth et al. (2012) are 
calculated from atmospheric models not only with H and He compositions,  
but also with C, N, O compositions, while the synthetic spectra in our 
study and Luo et al. (2016) are  calculated from  atmospheric models  
with only H and He composition (N\'emeth et al. 2014). 
A comparison of Fig 4 and Fig 5 reveals that at the given resolution and SNR the composition has only  a little influence on our final results.

\subsection{Parameter diagram}

\begin{figure}
\centering
\includegraphics [width=130mm]{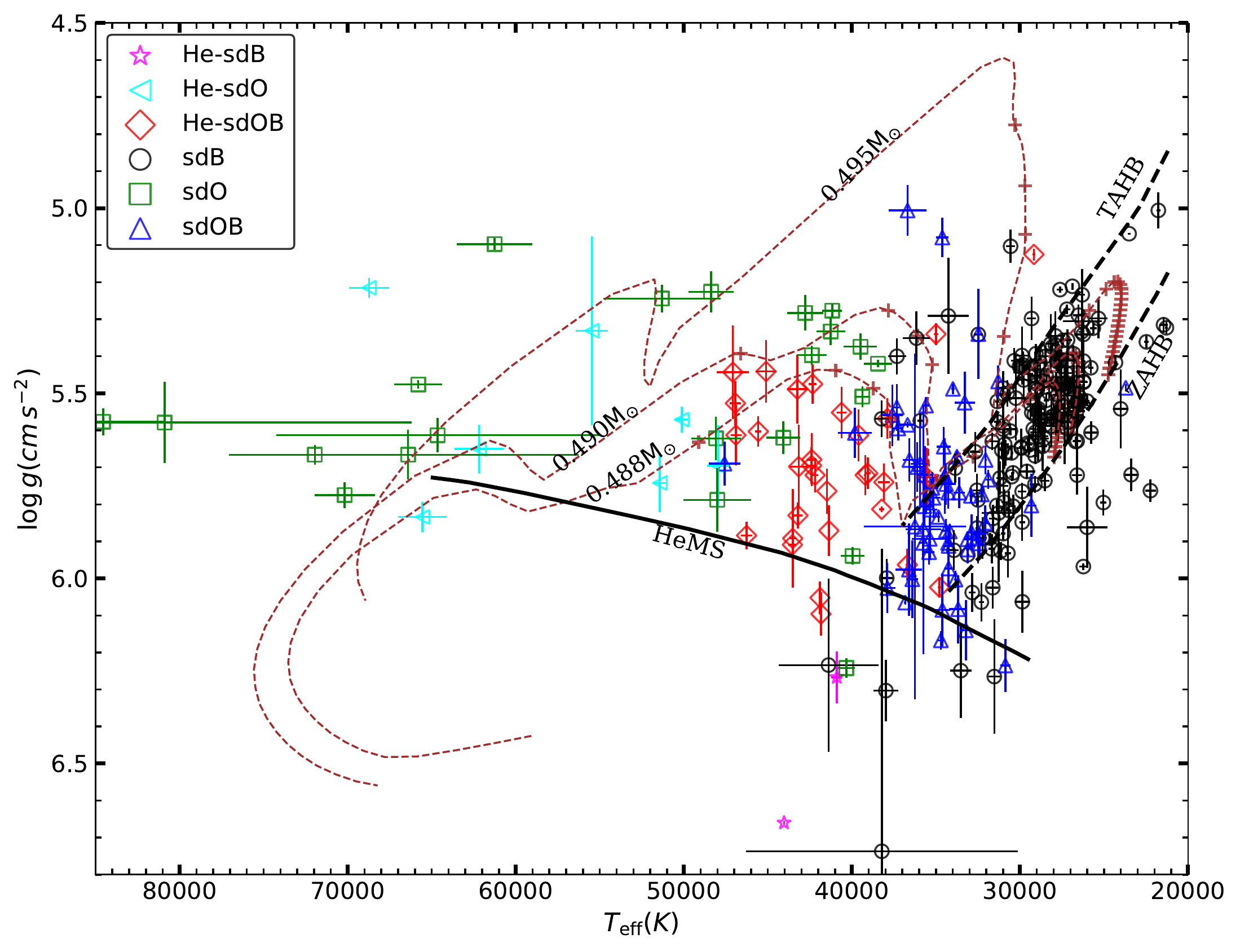}
\caption{$T_\mathrm{eff}$-$\mathrm{log}$ \textit{g} diagram 
for the 294 hot subdwarf stars identified in our study. 
He-sdB, He-sdO, He-sdOB, sdB,  sdO and sdOB stars are 
represented by magenta stars, aqua left triangles, red diamonds, 
black circles, green squares and blue up triangles, respectively. 
The thick solid line denotes the He-MS, which is taken from Paczy\'nski (1971), 
while the two dashed line represent the ZAHB and TAHB from Dorman et al. (1993) for [Fe/H]=-1.48. 
Three sdB evolution tracks from Dorman et al. (1993) for [Fe/H]=-1.48 are 
added, which are denoted by brown dashed lines. The stellar masses at the ZAHB are labeled close to each track. 
The time interval between two adjacent $+$ markers  is 5 Myr. }
\end{figure}

In Fig 6, we present the $T_\mathrm{eff}$-$\mathrm{log}\ g$ diagram for the identified 294  
hot subdwarf stars. The two dashed lines in the figure are 
the zero-age HB (ZAHB) and terminal-age HB (TAHB) calculated by  
Dorman et al. (1993) for [Fe/H]=-1.48, while the solid line denotes 
the He main-sequence (He-MS) from Paczy$\acute\mathrm{n}$ski (1971). 
The three brown dashed curves mark three sdB 
evolution tracks for [Fe/H]=-1.48, which are from Dorman et al. (1993), 
and the time interval between two adjacent $+$ marks  
is 5 Myr. The stellar masses at ZAHB are also labeled close to each track. 
He-sdB, He-sdO, He-sdOB, sdB,  sdO and sdOB stars are 
represented by different colors and marks, respectively. 

It is clear from Fig 6 that most of  the 
sdB stars (i.e., black circles) settle  in the region well defined by the ZAHB and 
TAHB lines, and they cluster near near $T_\mathrm{eff}=$ 29\,000 K 
and $\mathrm{log}\ g= 5.5\,cm\,s^{-2}$. Similarly, most of sdOB stars (i.e., blue up triangles) are located  
in the area between the ZAHB and TAHB as well, but they present a little higher 
temperatures and gravities than sdB stars, which cluster near the position of  
$T_\mathrm{eff}=$ 34\,000 K and $\mathrm{log}\ g= 5.8\,cm\,s^{-2}$. 
On the other hand, nearly all of the He-sdOB stars (red diamonds) are located above the 
TAHB and center at the position of $T_\mathrm{eff}=$ 40\,000 K and $\mathrm{log}\ g= 5.6\,cm\,s^{-2}$, 
while sdO stars (i.e., green squares)  and He-sdO (i.e., aqua left triangles) 
present a much more dispersive distribution 
in $T_\mathrm{eff}$-$\mathrm{log}\ g$ diagram  and higher 
temperatures than other type of hot subdwarf stars (e.g., some of 
them have $T_\mathrm{eff}>$ 60\,000 K). 

\begin{figure}
\centering
\includegraphics [width=120mm]{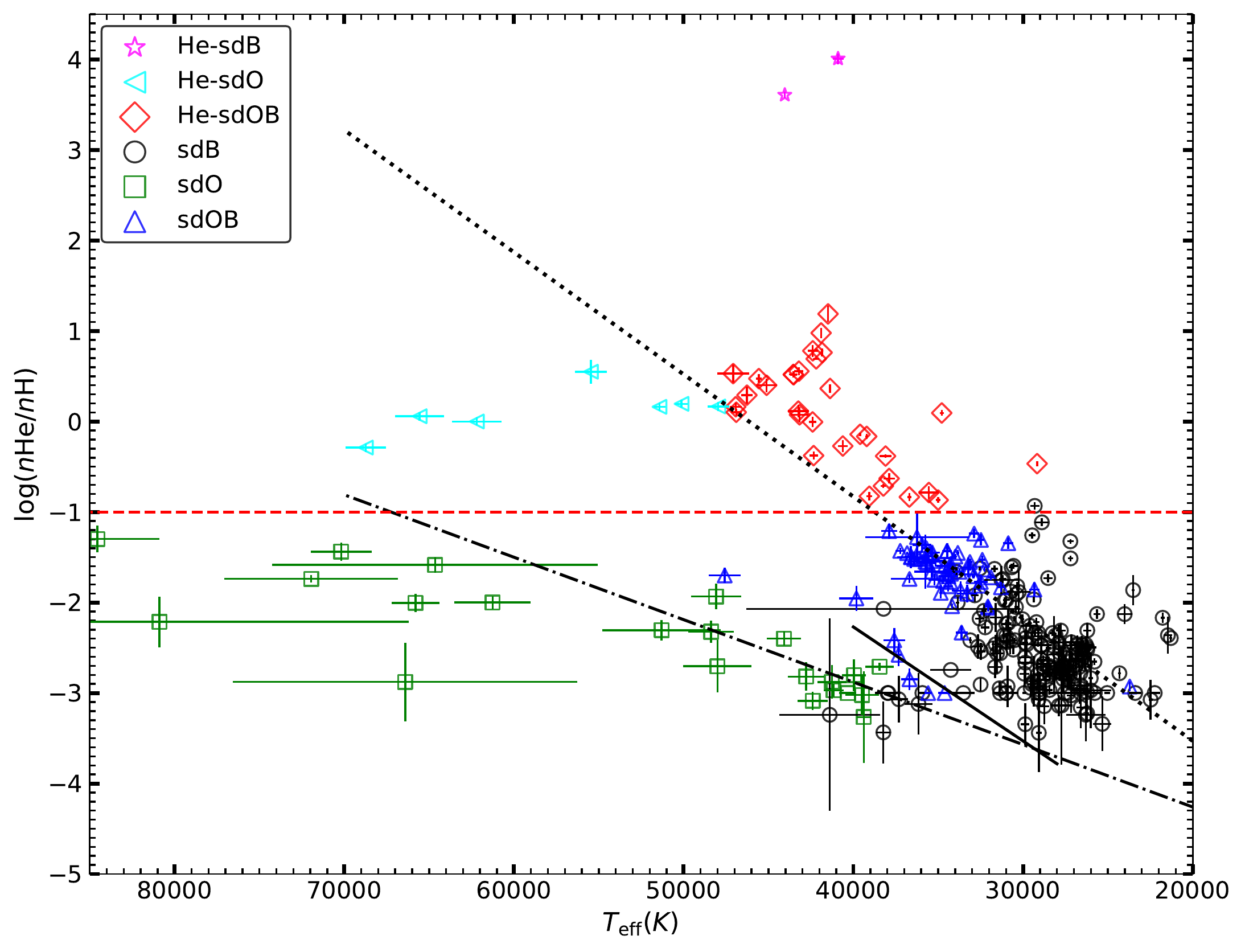}
\caption{$T_\mathrm{eff}$-$\mathrm{log}(n\mathrm{He}/n\mathrm{H})$ diagram for 
the 294 hot subdwarf stars identified in our study. 
Different types of hot subdwarf stars are denoted by the same markers as in 
Fig 6. The red dashed line denotes 
the solar He abundance. The dotted line and solid line are the linear regression line 
fitted by Edelmann et al. (2003), while the dot-dashed line is the best-fitting line 
for the He-weak sequence in N\'emeth et al. (2012).  }
\end{figure}

Fig 7 shows the $T_\mathrm{eff}$-$\mathrm{log}(n\mathrm{He}/n\mathrm{H})$ 
diagram  for the hot subdwarf stars in our sample. The different types  
of hot subdwarf stars are denoted by the same markers as in Fig 6. 
The red horizontal dashed line in this figure represents the solar He abundance 
(e.g., $\mathrm{log}(n\mathrm{He}/n\mathrm{H})=$ -1). 
As  described by other authors 
(e.g., Edelmann et al. 2003; N\'emeth et al. 2012; Geier et al. 2013; 
Luo et al. 2016), two distinct He sequences can be outlined in Fig 7 clearly. 
The black dotted line and black solid line are the linear regression lines found by  
Edelmann et al. (2003) to fit the He-rich and He-weak 
sequence in their study respectively, and they are described by the following 
equations:
\begin{equation}
\mathrm{log}(n\mathrm{He}/n\mathrm{H})=-3.53+1.35\left(\frac{T_\mathrm{eff}}{10^{4}K}-2.00\right), 
\end{equation}

\begin{equation}
\mathrm{log}(n\mathrm{He}/n\mathrm{H})=-4.79+1.26\left(\frac{T_\mathrm{eff}}{10^{4}K}-2.00\right). 
\end{equation}
 N\'emeth et al. (2012) used another linear regression line to 
fit the He-weak sequence in a larger parameter space, which is described by the following equation, 
\begin{equation}
\mathrm{log}(n\mathrm{He}/n\mathrm{H})=-4.26+0.69\left(\frac{T_\mathrm{eff}}{10^{4}K}-2.00\right). 
\end{equation}
This linear regression line is also presented by a black dot-dashed line in Fig 7. 
Fig 7 shows that both the He-rich and He-weak sequences 
show an increasing He abundance  with temperature. Moreover, 
our He-rich sequence consists of sdB (black circles), sdOB (blue triangles) 
, He-sdOB (red diamonds), He-sdO (aqua left triangles) and He-sdB (magenta) stars.  
Moreover, sdB stars have their He abundance 
in the rough range of -4 $<\mathrm{log}(n\mathrm{He}/n\mathrm{H})< $ -2, sdOB stars 
have their He abundance in the rough range of -2 $<\mathrm{log}(n\mathrm{He}/n\mathrm{H})< $ -1,  and 
most of He-sdOB stars are in the rough He abundance range of 
-1 $<\mathrm{log}(n\mathrm{He}/n\mathrm{H})<$ 2.  
He-sdO stars have similar He abundances as He-sdOB stars in our sample, but they present 
higher temperatures than He-sdOB stars. We have 2 He-sdB stars (magenta stars), which 
present the highest He abundance  in our sample. 
Our He-rich sequence could be  approximately fitted by the line found by Edelmann et al. (2003). 
The He-weak sequence consists of all sdO stars 
and a few of sdB  and sdOB stars with very low He abundance. 
Furthermore, the He-weak sequence  is quite diverse and better reproduced  by the fitting line used in  
N\'emeth et al. (2012) than the one used in Edelmann et al. (2003). 
\begin{figure}
\centering
\includegraphics [width=120mm]{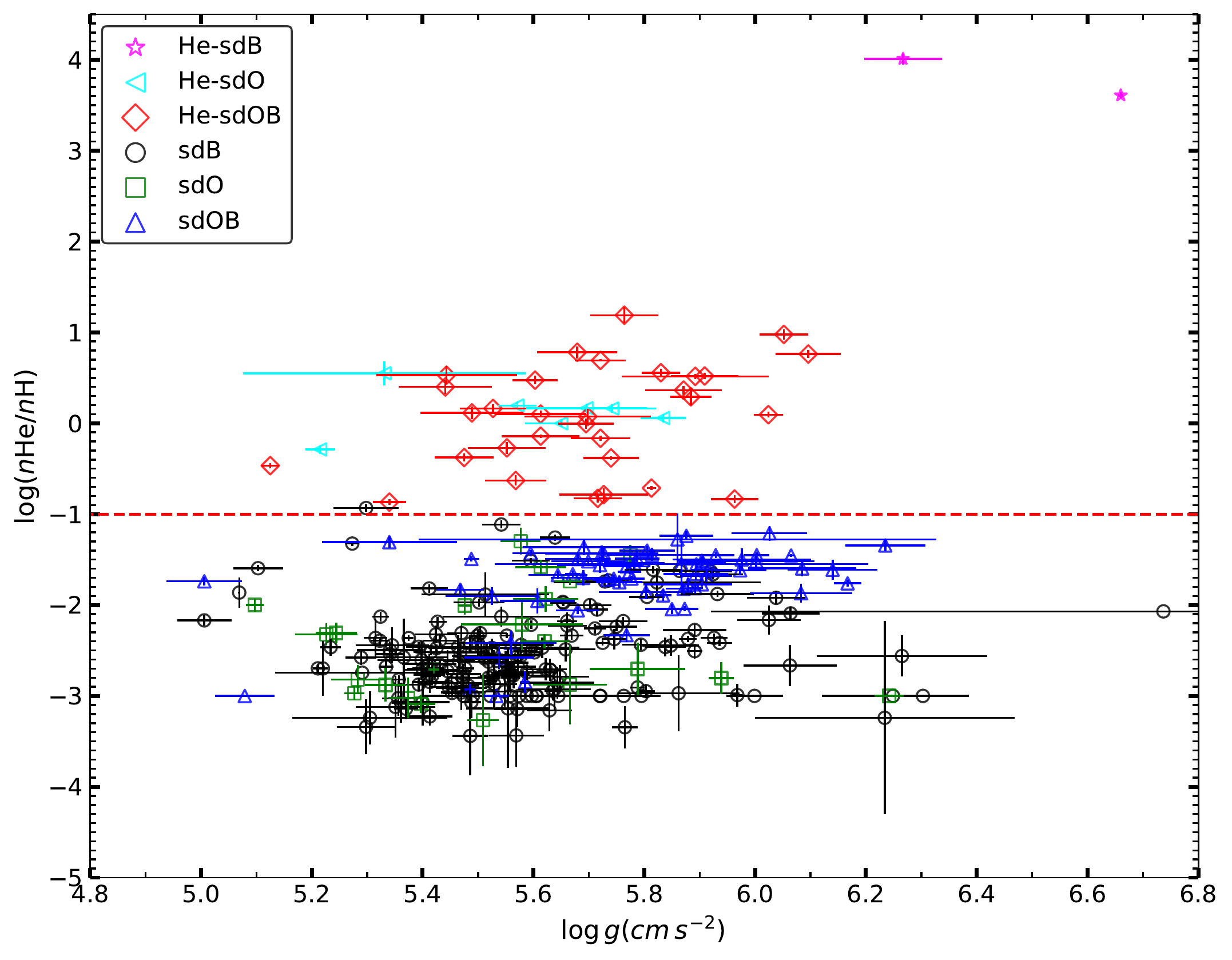}
\caption{$\mathrm{log}$ \textit{g}-$\mathrm{log}(n\mathrm{He}/n\mathrm{H})$ diagram for the 294 
hot subdwarf stars identified in our study.  The red dashed line denotes the solar He abundance. Different 
types of hot subdwarf stars are denoted by the same markers as in Fig 6. }
\end{figure}

We show the $\mathrm{log}\ g$-$\mathrm{log}(n\mathrm{He}/n\mathrm{H})$ 
diagram for our identified hot subdwarf stars in Fig 8. Like in  
Fig 6 and 7, different types of hot subdwarf stars are denoted 
by different markers, and the red dashed line is the solar 
He abundance (e.g., $\mathrm{log}(n\mathrm{He}/n\mathrm{H})=$ -1). 
In this figure, no distinct sequences are presented and, in general,  
all hot subdwarf stars independent of their spectral  types show a 
wide distribution of surface gravities. 

\section{discussion and summary }

\begin{figure}
\centering
\begin{minipage}[c]{0.45\textwidth}
\includegraphics [width=70mm]{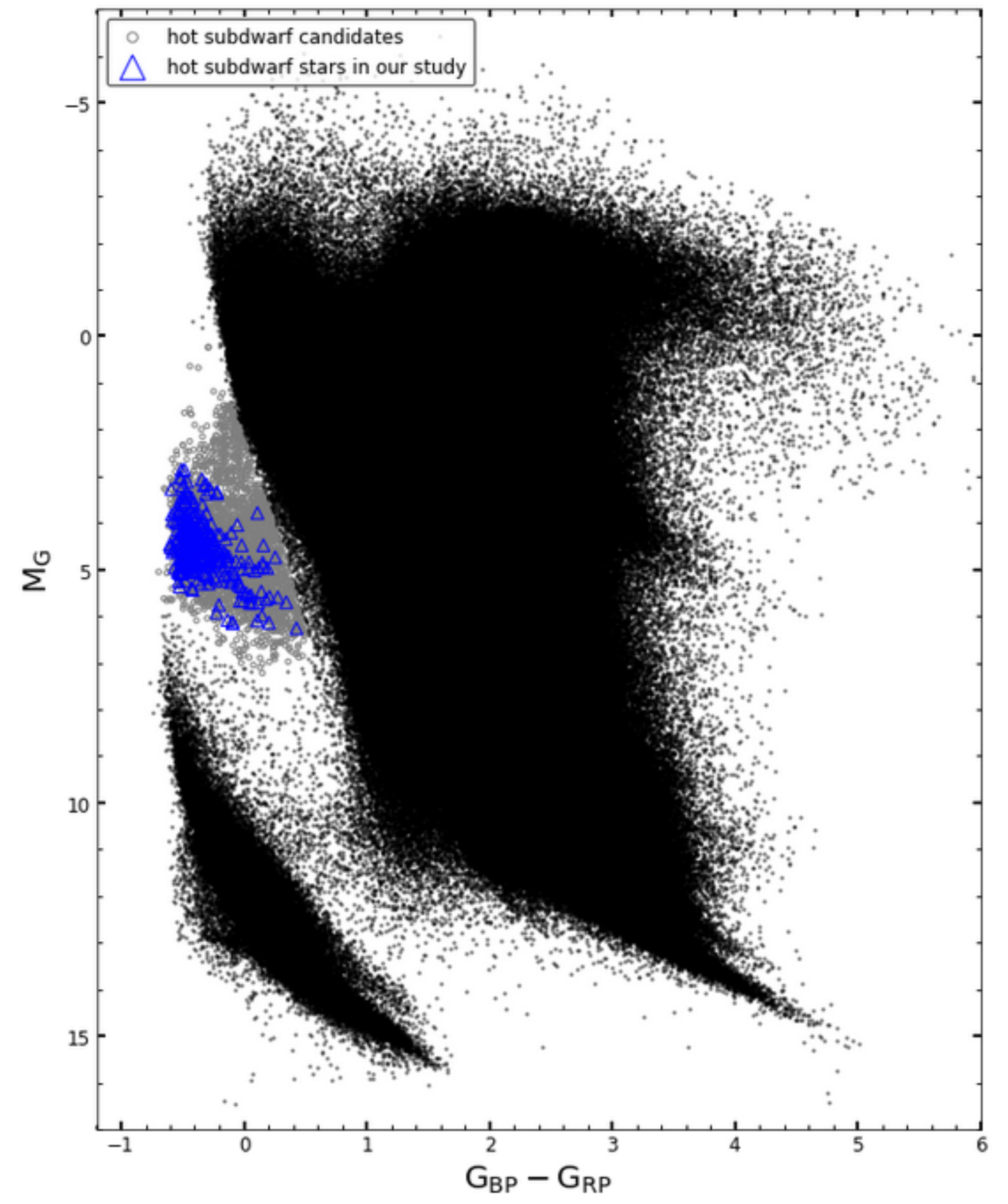}
\centerline{(a) }
\end{minipage}%
\begin{minipage}[c]{0.45\textwidth}
\includegraphics [width=70mm]{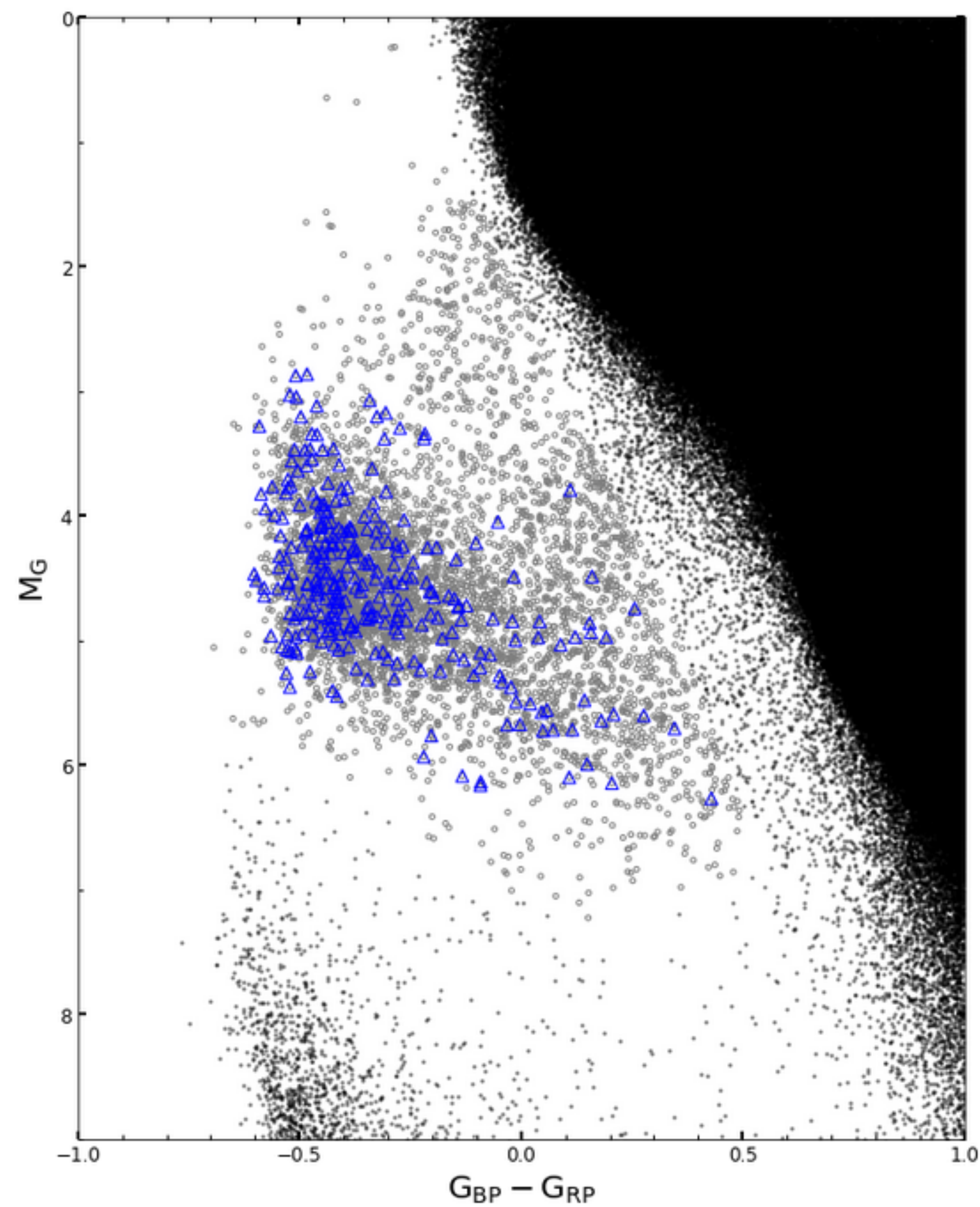}
\centerline{(b) }
\end{minipage}%

\caption{Panel (a) : Gaia HR diagram. Hot subdwarf candidates (e.g., 4593 objects) are marked with  
gray circles. 294 hot subdwarf  stars confirmed in our study are marked with  
blue triangles. Panel (b) : Magnified drawing for hot subdwarf region. }
\end{figure}

With the accurate astrometry and photometry from Gaia DR2, it is now possible 
to built a HR diagram with huge number of homogeneous data (Gaia collaboration et al. 2018b), 
which eventually provide us a new tool to study stellar evolution for 
the stars at different evolutionary stages. 
Panel (a) of Fig 9 shows the hot subdwarf stars  identified in  
the Gaia DR2 HR diagram (Gaia collaboration et al. 2018b, see Section 2.1 for details). 
The gray dots within the polygon are the hot subdwarf candidates 
(i.e., 4593 stars) selected in our study (see Section 2), 
while the blue triangles are the hot subdwarf stars identified in our study. 
Panel (b) of Fig 9 is the magnified region where 
we selected the hot subdwarf candidates. 

Fig 9 shows clearly that the majority of  hot subdwarf stars are found  
at the typical position where they are expected in the HR diagram, e.g., clustering at 
$\mathrm{M}_{G}\thickapprox$ 5 and BP-RP$\thickapprox$ -0.2. 
However some stars with a  spectroscopic subdwarf confirmation 
appear  far from the subdwarf region, e.g., 
a few of stars extend the hot subdwarf distribution to the right and 
connect it with the wide MS. Some of these stars 
are hot subdwarf binaries with a WD or low mass MS 
companion (a detailed study for the hot subdwarf binaries 
are out of the scope of this work), while some spectra may suffer from the effects of unknown 
extinction (Andrae et al. 2018) and present  redder positions in the  HR diagram than 
normal hot subdwarf stars. 

We have selected 4593 hot subdwarf candidates by means of the Gaia DR2 HR diagram, among which more 
than 700 objects have spectra in LAMOST DR5. Finally, 
we confirmed 294 hot subdwarf stars from our candidates, 
including 169 sdB, 63 sdOB, 31 He-sdOB, 22 sdO, 
7 He-sdO and 2 He-sdB stars. 
Considering hot subdwarfs that are not listed in Geier et al 2017 as new discoveries, 
we found 110 new hot subdwarfs in our study. Note however, 
that besides the  hot subdwarf candidates which have 
LAMOST spectra, there are more 
than 3800 candidates without spectra in LAMOST, and most of them are likely real hot subdwarf stars 
based on their positions near the hot subdwarf region in Gaia HR diagram. 
In future work we will use LAMOST to obtain the spectra of these candidates, which 
will eventually help us obtain their atmospheric parameters and spectral classification types. On the other hand, 
the samples  which were used to built the HR diagram in Gaia collaboration et al. (2018b) 
have only the most precise parallaxes and photometry. These samples represent only the tip of the iceberg. 
We hope to find many more hot subdwarf stars by combining the LAMOST spectroscopic database with the whole Gaia DR2 source list.

\acknowledgments 
This work is supported by the National Natural Science Foundation 
of China Grant Nos 11390371, 11503016 and U1731111,   
Natural Science Foundation of Hunan province Grant No.2017JJ3283,  
the Youth Fund project of Hunan Provincial Education Department
Grant No.15B214, the Astronomical Big Data Joint Research
Center, co-founded by the National Astronomical Observatories, Chinese 
Academy of Sciences and the Alibaba Cloud. 
This research has used the services of \mbox{\url{www.Astroserver.org}} under reference X074VU and KS5NVO. 
P.N. acknowledges support from the Grant Agency of the Czech Republic (GA\v{C}R 18-20083S). 
The LAMOST Fellowship is supported by Special Funding for Advanced Users, 
budgeted and administered by the Center for Astronomical 
Mega-Science, Chinese Academy of Sciences (CAMS). 
Guoshoujing Telescope (the Large Sky Area Multi-Object Fiber 
Spectroscopic Telescope LAMOST) is a National Major Scientific 
Project built by the Chinese Academy of Sciences. 
Funding for the project has been provided by the 
National Development and Reform Commission. 
LAMOST is operated and managed by the National Astronomical Observatories, 
Chinese Academy of Sciences.



\end{document}